\documentclass[%
 prd,
 twocolumn,
 superscriptaddress,
 numerical,
 showpacs,
 amsmath,amssymb,
 aps,
 nofootinbib,
 longbibliography,
 floatfix,
 10pt
]{revtex4-1}

\usepackage{bm}
\usepackage{bbm}
\usepackage{soul}
\usepackage{cancel}
\usepackage{braket}
\usepackage{slashed}
\usepackage{verbatim}
\usepackage{graphicx}
\usepackage{hyperref}
\usepackage{combelow} 
\usepackage{mathtools}
\usepackage[dvipsnames,svgnames,table]{xcolor}

\hypersetup{
  colorlinks,
  citecolor=Violet,
  linkcolor=Blue,
  urlcolor=teal}

\let\oldciteauthor=\citeauthor
\def\citeauthor#1{\hypersetup{citecolor=black}\oldciteauthor{#1}}

\let\oldciten=\onlinecite
\def\onlinecite#1{\hypersetup{citecolor=blue}\oldciten{#1}}

\let\oldcite=\cite
\def\cite#1{\hypersetup{citecolor=blue}\oldcite{#1}}

\allowdisplaybreaks

\begin{document}

\title{Strongly interacting fermions under imaginary rotation}

\author{Tudor P\u{a}tuleanu}
\affiliation{Department of Physics, West University of Timi\cb{s}oara,  Bd.~Vasile P\^arvan 4, Timi\cb{s}oara 300223, Romania}

\affiliation{Institut Denis Poisson, CNRS UMR 7013, Universit\'e de Tours, Universit\'e d'Orl\'eans, 
Parc de Grandmont, Tours, 37200, France}


\author{Victor E. Ambru\cb{s}}
\thanks{Corresponding author: victor.ambrus@e-uvt.ro.}
\affiliation{Department of Physics, West University of Timi\cb{s}oara,  Bd.~Vasile P\^arvan 4, Timi\cb{s}oara 300223, Romania}

\begin{abstract}
In the present study, we investigate the phase diagram of strongly-interacting fermions under imaginary rotation. We employ the linear sigma model coupled to quarks in the mean field approximation. The fermion expectation values are computed in the local density approximation (LDA), using the well-known expressions derived using cylindrical modes for states under rigid rotation, at finite temperature $T$ and chemical potential $\mu$. We study the impact of the fractalization of thermodynamics on the phase diagram in the far-field limit (far from the rotation axis). We also reveal non-trivial features such as a complete inhibition of the chiral symmetry restoration on the rotation axis above a critical imaginary angular velocity. We demonstrate explicitly how the transition line interpolates between its shape on the rotation axis and in the far-field limit by studying the chiral phase transition at finite distances from the rotation axis. We also address the moment of inertia of the system and discuss differences and similarities to lattice studies of QCD matter under imaginary rotation.
\end{abstract}

\maketitle

\section{Introduction}\label{sec:intro}

The impact of rotation on the phase diagram of strongly interacting matter has seen a significant increase of interest over the past decade or so. Owing to the groundbreaking measurement \cite{STAR:2017ckg,STAR:2018gyt} by the STAR (Solenoidal Tracker At RHIC) collaboration at RHIC (Relativistic Heavy-Ion Collider) in BNL (Brookhaven National Laboratory) of the persistent polarization of Lambda hyperons in heavy-ion collisions, the angular velocity of the rotation of the quark-gluon plasma was estimated at $\Omega \sim 6.6$ MeV \cite{STAR:2017ckg}. As the state of rigid rotation can be ascribed to a Killing vector of the Minkowski space-time, such states can be in global thermal equilibrium \cite{Cercignani.2002,Becattini:2012tc}, provided suitable boundary conditions are enforced to prevent the violation of causality \cite{Nicolaevici:2001yy,Duffy:2002ss,Ambrus:2015lfr}. A natural question arises regarding the impact of vorticity on the details of the transition of strongly interacting matter, described by quantum chromodynamics (QCD), from the deconfined, chirally restored quark-gluon plasma (QGP) phase to the hadronic phase.

The study of the phase transition in QCD can be done from first principles using the lattice QCD (lQCD) technique, which relies on the numerical evaluation of the QCD path integral using Monte Carlo techniques. Due to the infamous sign problem, which traditionally plagues finite chemical potential calculations, calculations for systems under rotation must be performed using an imaginary rotation angular velocity, $\Omega = i \Omega_I$ (with $\Omega_I$ a real number) \cite{Yamamoto:2013zwa}. Such studies reveal certain nonconventional features of the rotating quark-gluon plasma. 

First, the Tolman-Ehrenfest law implies that the local temperature $T_\rho$ inside the rigidly rotating plasma increases with the distance $\rho$ from the rotation axis, via the formula \cite{Cercignani.2002}
\begin{equation}
 T_\rho = T \Gamma_\rho, \quad 
 \Gamma_\rho = \frac{1}{\sqrt{1 - \rho^2 \Omega^2}},
 \label{eq:Trho}
\end{equation}
with $T$ being the temperature on the rotation axis and $\Gamma_\rho$ the Lorentz factor of an observer on a co-rotating circular trajectory of radius $\rho$. When $\Omega = i \Omega_I$, $\Gamma_\rho \to \Gamma_{\rho;I} = 1/\sqrt{1 + \rho^2 \Omega_I^2}$ and Eq.~\eqref{eq:Trho} predicts a {\it decrease} of the local temperature with the distance $\rho$. Hence, the system should be getting {\it colder}, thus inhibiting the transition to the QGP. However, lQCD results indicate a contrary trend: the system set up in the confined phase close to the rotation axis becomes deconfined far from the rotation axis \cite{Braguta:2023yjn,Braguta:2023tqz,Braguta:2024zpi,Yang:2023vsw}. 

The second remarkable feature uncovered by lQCD calculations is that the moment of inertia $\mathcal{I}$ takes a negative value above the deconfinement temperature $T_{\rm dec}$, up to the so-called supervortical temperature $T_{\rm sv} \simeq 1.5 T_{\rm dec}$ \cite{Braguta:2023yjn, Braguta:2024zpi, Braguta:2023tqz}. The lQCD simulations imply that this behavior is due to the kinetic term in the gluon Lagrangian, associated with the chromomagnetic part of the field strength tensor, and not due to the fermion contribution \cite{Braguta:2024zpi}. 

Both features have proven difficult to reproduce using traditional effective models of QCD. Since the sign problem does not affect calculations in such models, previous investigations have considered the effect of real rotational angular velocity. Both the Nambu--Jona-Lasinio model (NJL) \cite{Chernodub:2016kxh,Wang:2018sur} and the linear sigma model coupled with quarks (LSM$_q$) \cite{Sun:2023kuu,Singha:2024tpo,Singha:2025zvh}, as well as its functional renormalization group version \cite{Chen:2023cjt} give results consistent with the Tolman-Ehrenfest prediction \cite{Singha:2025zvh}, and in contradiction with lQCD results. Other studies, involving 2+1-dimensional (2+1-D) electrodynamics \cite{Chernodub:2020qah}, perturbative Yang-Mills \cite{Chen:2022smf} and perturbative QCD \cite{Fujimoto:2021xix,Chen:2024tkr}, are also in disagreement with lQCD data. The inclusion of nonperturbative effects in SU(2) Yang-Mills \cite{Jiang:2024zsw} seems to restore agreement in respect to the radial structure of the inhomogeneous phase, but cannot recover the negative moment of inertia.  Note that agreement with lattice data can be restored in effective models where the coupling parameters are promoted to rotation-dependent functions, such as the NJL model \cite{Jiang:2021izj} and its Polyakov-enhanced version \cite{Sun:2024anu} or SU(2) Yang-Mills \cite{Jiang:2023zzu}.

In this work, we study the properties of the chiral phase transition at imaginary rotation using the linear sigma model coupled to quarks (LSM$_q$). Starting from our previous work \cite{Patuleanu:2025zbn}, where we studied free fermions under imaginary rotation, we investigate the thermodynamics of the LSM$_q$ model by constructing the thermodynamic potential density $\phi(x)$ at finite fermion mass. We employ the mean-field approximation, by which the $\sigma$ and $\vec{\pi}$ mesons are treated as classical fields, whose expectation values are given by locally minimizing the total (fermion $+$ meson) thermodynamic potential density. We further work in the local density approximation (LDA), evaluating the thermodynamic potential at each point with the local value of the fermion effective mass $M = g \sigma$ as if it were constant throughout the whole space. A more elaborate approach going beyond the LDA, which consistently accounts for the inhomogeneity of the meson field, involves solving the Bogolubov-de Gennes equation \cite{Wang:2019nhd,Wang:2018zrn}, however we do not pursue such extensions in this work. We further neglect the spatial gradients of the meson fields when evaluating the full thermodynamic potential. A discussion on the role of these gradients for a system under real rotation can be found in Ref.~\cite{Sergio:2025zrn}. 

To motivate our analysis, we first remind that the lQCD studies of systems under rotation are performed at imaginary rotation. As shown in Refs.~\cite{Ambrus:2023bid} and \cite{Patuleanu:2025zbn} for the scalar and Dirac fields, respectively, imaginary rotation introduces nontrivial and unexpected features, which must be taken into account before considering the analytic continuation of results obtained by such studies to the case of real rotation angular velocity. 

The most striking feature is the fractalization of thermodynamics in the far-field limit (far from the rotation axis), where the system resembles a static system with reduced temperature $T_\mathsf{q} = T/\mathsf{q}$ and the same chemical potential $\mu$, where $\mathsf{q}$ is the denominator of the irreducible fraction representation of the dimensionless rotation parameter, $\nu = \beta \Omega_I / 2\pi = \mathsf{p} / \mathsf{q}$. This is true for the scalar field, studied in Ref.~\cite{Ambrus:2023bid}, for all values of $\mathsf{p}$ and $\mathsf{q}$, while in the case of the Dirac field, the system remains fermionic only when $\mathsf{k} = \mathsf{p} + \mathsf{q}$ is an odd integer. This is related to a second remarkable feature, the emergence of so-called ninionic statistics introduced by Chernodub \cite{Chernodub:2022qlz}. Depending on the angular momentum of the mode, the effective statistics can be either bosonic, fermionic or ghost-like \cite{Chernodub:2020qah}. A striking consequence for fermions is that the ninionic statistics persist in the far-field limit, with the fermionic (ghost) behavior achieved when $\mathsf{k} = \mathsf{p} + \mathsf{q}$ is an odd (even) integer. A third feature is that certain quantities, such as the energy-momentum tensor or the fermion condensate, can become negative at large $\nu$, even on the rotation axis. All of these features have significant (and non-trivial) consequences on the chiral restoration phase diagram of the system, as will be discussed in this work. Despite these exotic features, our results for the effect of imaginary rotation on the chiral restoration transition remain in disagreement with lattice results.

The outline of this paper is as follows. In Sec.~\ref{sec:LSMq}, we review the LSM$_q$ model and discuss its formulation under rotation. In Sec.~\ref{sec:rot}, we discuss the quark contribution to the thermodynamic potential and evaluate it in the case of rational rotation parameter $\nu = \beta \Omega_I / 2\pi= \mathsf{p}/\mathsf{q}$, where $\mathsf{p}$ and $\mathsf{q}$ are coprime integers. The phase diagram of the system is discussed in Sec.~\ref{sec:pd}, with special emphasis on the far-field limit (Sec.~\ref{sec:pd:far}), rotation axis (Sec.~\ref{sec:pd:axis}) and finite distances (Sec.~\ref{sec:pd:l}). The moment of inertia of the system is discussed in Sec.~\ref{sec:I}. Conclusions are presented in Sec.~\ref{sec:conc}. 

\section{\texorpdfstring{LSM$_q$}{LSMq} model under rotation}\label{sec:LSMq}

In this section, we introduce the linear sigma model coupled to quarks (LSM$_q$) \cite{Gell-Mann:1960mvl,Scavenius:2000qd,Singha:2025zvh} at finite temperature, chemical potential and rotation angular velocity. Section~\ref{sec:LSMq:L} presents the LSM$_q$ Lagrangian and the model parameters. Section~\ref{sec:LSMq:mean} introduces the mean-field approximation and the saddle-point equation governing the expectation value of the $\sigma$ and $\vec{\pi}$ mesons. Section~\ref{sec:LSMq:rot} presents the formulation of the model under rotation in the local density approximation (LDA).

\subsection{\texorpdfstring{LSM$_q$}{LSMq} Lagrangian} \label{sec:LSMq:L}

The linear sigma model coupled to quarks (LSM$_q$) is described by the Lagrangian density
\begin{equation}
 \mathcal{L} = \mathcal{L}_m + \mathcal{L}_\psi,
\end{equation}
where $\mathcal{L}_m$ describing the meson sector $(\sigma, \vec{\pi})$ reads
\begin{equation}
 \mathcal{L}_m = \frac{1}{2} \partial_\mu \sigma \partial^\mu \sigma + \frac{1}{2} \partial_\mu \vec{\pi} \cdot \partial^\mu \vec{\pi} - V_m.
\end{equation}

Besides the kinetic terms for the scalar ($\sigma$) and pseudoscalar triplet ($\vec{\pi}$) mesons, the model includes a quartic potential,
\begin{equation}
 V_m = \frac{\lambda}{4}(\sigma^2 + \vec{\pi}^2 - v^2)^2 - h \sigma,
\end{equation}
where the constant parameters $\lambda$, $v^2$ and $h$ are fixed by the vacuum properties of the theory. In the vacuum, the meson fields take the expectation values $\sigma_0$ and $\vec{\pi}_0$ that minimize $V_m$, namely
\begin{align}
 \frac{dV_m}{d\sigma} &= \lambda(\sigma_0^2 + \vec{\pi}_0^2 - v^2) \sigma_0 - h = 0, \nonumber\\
 \frac{dV_m}{d\vec{\pi}} &= \lambda(\sigma_0^2 + \vec{\pi}_0^2 - v^2) \vec{\pi}_0 = 0.
\end{align}

This implies $\vec{\pi}_0 = 0$ and $\lambda(\sigma_0^2 - v^2) \sigma_0 = h$. Furthermore, the vacuum meson masses evaluate to 
\begin{align}
 m_\sigma^2 &= \frac{d^2 V_m}{d\sigma^2} = \lambda(3\sigma_0^2 - v^2), \nonumber\\
 m_\pi^2 &= \frac{d^2 V_m}{d\vec{\pi}^2} = \lambda(\sigma_0^2 - v^2).
\end{align}

Imposing $\sigma_0 = f_\pi = 93$ MeV, where $f_\pi$ is the pion decay constant, as well as $m_\sigma = 600$ MeV and $m_\pi = 138$ MeV, gives 
\begin{equation}
 \lambda = 19.71, \quad 
 v = 87.7\ {\rm MeV}, \quad 
 h = (121\ {\rm MeV})^3.
\end{equation}

For the quark part, we consider $N_f = 2$ massless flavors (the $u$ and $d$ flavors), with a color degeneracy factor of $N_c = 3$. The corresponding Lagrangian reads
\begin{equation}
 \mathcal{L}_\psi = \bar{\psi} [i \slashed{\partial} - g(\sigma + i \gamma^5 \vec{\tau} \cdot \vec{\pi})] \psi,
\end{equation}
where the spinor $\psi$ collects all $N_f \times N_c = 2 \times 3 = 6$ quark species, each represented by a four-component spinor. In the above, $\vec{\tau}$ are the usual $SU(2)$ Pauli matrices in flavor space, $\slashed{\partial} = \gamma^\mu \partial_\mu$ and the Dirac matrices $\gamma^\mu$, satisfying $\{\gamma^\mu, \gamma^\nu\} = \gamma^\mu \gamma^\nu + \gamma^\nu \gamma^\mu = 2 g^{\mu\nu}$, are taken in the Dirac representation:
\begin{equation}
 \gamma^0 = \begin{pmatrix}
     1 & 0 \\ 0 & -1
 \end{pmatrix}, \quad 
 \gamma^i =  \begin{pmatrix}
      0& \sigma^i \\ 
      -\sigma^i & 0
 \end{pmatrix},
\end{equation}
with the Pauli matrices given by
\begin{equation}
 \sigma^1 = \begin{pmatrix}
    0 & 1 \\ 1 & 0
 \end{pmatrix}, \quad 
 \sigma^2 = \begin{pmatrix}
     0 & -i\\ i & 0
 \end{pmatrix}, \quad 
 \sigma^3 = \begin{pmatrix}
     1 &0 \\ 0 & -1
 \end{pmatrix}.
\end{equation}
The coupling constant $g$ is chosen such that the effective mass of the quarks in vacuum is $M_0 = g \sigma_0 = 307$ MeV, i.e. $g = 3.3$.

The meson and quark fields satisfy the Euler-Lagrange equations of motion
\begin{align}
 \left[\Box + \lambda (\sigma^2 + \vec{\pi}^2 - v^2)\right] \sigma &= h - g \bar{\psi} \psi, \nonumber\\
 \left[\Box + \lambda (\sigma^2 + \vec{\pi}^2 - v^2)\right] \vec{\pi} &= - ig \bar{\psi} \gamma^5 \vec{\tau} \psi, \nonumber\\
 [i \slashed{\partial} - g (\sigma + i \gamma^5 \vec{\tau} \cdot \vec{\pi})] \psi &= 0.
\end{align}

\subsection{Mean field approximation} \label{sec:LSMq:mean}

In the mean field approximation, the meson fields are treated as classical fields. Furthermore, in this work, we consider that their expectation values vary slowly throughout the domain, such that the kinetic terms $\Box \sigma$ and $\Box \vec{\pi}$ are neglected. Moreover, as we will see shortly, the expectation value $\langle i \bar{\psi} \gamma^5 \vec{\tau} \psi \rangle$ of the pseudoscalar condensate vanishes for the states that we consider [see above Eq.~\eqref{eq:SC_gen}]. For this reason, the pion field takes a vanishing expectation value, $\vec{\pi} = 0$, and the sigma field is found by solving
\begin{equation}
 \lambda(\sigma^2 - v^2) \sigma = h - g \langle \bar{\psi} \psi \rangle,
 \label{eq:sigma_eq}
\end{equation}
while the Dirac field satisfies
\begin{equation}
 (i \slashed{\partial} - g \sigma) \psi = 0.
 \label{eq:Dirac_eq}
\end{equation}

Since, under rigid rotation, the scalar condensate $\langle \bar{\psi} \psi\rangle$ is inhomogeneous, the sigma field also depends on the radial distance $\rho$ to the rotation axis, $\sigma \equiv \sigma(\rho)$. Under the local density approximation, we neglect the variation of $\sigma$ with respect to the radial coordinate when solving the Dirac Eq.~\eqref{eq:Dirac_eq}. In the LDA, when evaluating $\sigma(\rho_0)$ at the point $\rho_0$, we consider solutions of Eq.~\eqref{eq:Dirac_eq} corresponding to the effective mass $M(\rho_0) = g \sigma(\rho_0)$, i.e.
\begin{equation}
 [i \slashed{\partial} - M(\rho_0)] \psi = 0.
\end{equation}

\subsection{Model under rotation} \label{sec:LSMq:rot}

In order to evaluate the scalar condensate $\langle \bar{\psi} \psi \rangle$ at distance $\rho_0$ from the rotation axis, we employ the density operator 
\begin{equation}
 \hat{\rho} = e^{-\beta (\widehat{H} - \mu \widehat{Q} - \Omega \widehat{M}^z)},
\end{equation}
where $\widehat{H}$, $\widehat{Q}$ and $\widehat{M}^z$ are the Hamiltonian, charge and angular momentum operators, respectively. Here, $\beta = 1/T$ and $\mu$ represent the inverse temperature and chemical potential on the rotation axis, while $\Omega$ is the constant angular velocity of rotation about the $z$ axis. 

The expectation value for the pseudoscalar and scalar condensates of fermions of constant mass $M = g \sigma$\footnote{Under the LDA, we compute fermion expectation values with the local value of the $\sigma$ meson as if it where constant throughout space.} under rotation was computed in Ref.~\cite{Ambrus:2019ayb}, being given by $\langle i \bar{\psi} \gamma^5 \vec{\tau} \psi \rangle = 0$ and
\begin{multline}
 \langle \bar{\psi} \psi \rangle = 
 \frac{M N_c N_f}{4\pi^2} \sum_{\varsigma = \pm 1} \sum_{m = -\infty}^\infty \int_M^\infty \frac{dE}{e^{\beta \widetilde{\mathcal{E}}_\varsigma} + 1} \\
 \times \int_{-p}^p dk\, J_m^+(q\rho), \label{eq:SC_gen}
\end{multline}
where $m = \pm \frac{1}{2}, \pm \frac{3}{2}, \dots$ is the projection of the total angular momentum on the $z$ axis, $\varsigma = \pm 1$ distinguishes between particle and anti-particle contributions, $\widetilde{\mathcal{E}}_\varsigma = \mathcal{E}_\varsigma - \Omega m$ is the corotating effective energy, with $\mathcal{E}_\varsigma = E - \varsigma \mu$, $k$ is the vertical component of the linear momentum, $p = \sqrt{E^2 - M^2}$ is the momentum magnitude and $q = \sqrt{p^2 - k^2}$ is the transverse momentum magnitude. In the above, we introduced the notation
\begin{equation}
 J_m^+(q\rho) = J_{m - \frac{1}{2}}^2(q\rho) + J_{m + \frac{1}{2}}^2(q\rho),
\end{equation}
with $J_n(z)$ being the Bessel function of the first kind. 

Equation~\eqref{eq:sigma_eq} can be understood as the saddle point equation $d\phi / d\sigma = 0$, extremizing the thermodynamic (grand) potential density $\phi(x) = d\Phi / dV$, with $\Phi = \int d^3x\, \phi(x)$, as detailed in Ref.~\cite{Sergio:2025zrn}. In the vicinity of a first-order transition, Eq.~\eqref{eq:sigma_eq} admits multiple solutions, out of which we must choose the one that minimizes $\phi$, which takes the form
\begin{equation}
 \phi(x) = V_\sigma(x) + \phi_\psi(x),
\end{equation}
where $V_\sigma(x) = V_m\rvert_{\vec{\pi} = 0} = \frac{\lambda}{4}(\sigma^2 - v^2)^2 - h \sigma$ is evaluated using the local value of $\sigma$, while the quark (fermion) part reads \cite{Jiang:2016wvv,Fujimoto:2021xix,Ambrus:2025dca}
\begin{multline}
 \phi_\psi = -\frac{N_c N_f T}{4\pi^2} \sum_{\varsigma = \pm 1} \sum_{m = -\infty}^\infty \int_M^\infty dE\, E\, \ln (1 + e^{-\beta \widetilde{\mathcal{E}}_\varsigma}) \\\times 
 \int_{-p}^p dk\, J_m^+(q\rho),
 \label{eq:phi_ln}
\end{multline}
satisfying $d\phi_\psi / dM = \langle \bar{\psi} \psi \rangle$. Using an integration by parts, the thermodynamic potential can be put in the following form:
\begin{multline}
  \phi_\psi = -\frac{N_c N_f}{4\pi^2} \sum_{\varsigma = \pm 1} \sum_{m = -\infty}^\infty \int_M^\infty \frac{dE}{e^{\beta \widetilde{\mathcal{E}}_\varsigma} + 1} \\\times 
 \int_{-p}^p dk\, k^2 J_m^+(q\rho).
 \label{eq:phi_gen}
\end{multline}

\section{Massive fermions under imaginary rotation} \label{sec:rot}

In this section, we evaluate the fermion condensate and the quark potential under imaginary rotation, $\Omega = i \Omega_I$ with $\Omega_I \in \mathbb{R}$ a real number, following the procedure outlined in Ref.~\cite{Patuleanu:2025zbn}. We begin with a discussion valid for arbitrary values of the dimensionless rotation parameter $\nu = \beta \Omega_I/ 2\pi$ in Sec.~\ref{sec:rot:arbitrary}. The case of rational rotation parameter, represented as the irreducible fraction $\nu = \mathsf{p} / \mathsf{q}$, is discussed in Sec.~\ref{sec:rot:rational}, where we address the emergence of fractal thermodynamics. We discuss the thermodynamics of the state (entropy, angular momentum, charge and energy densities) in Sec.~\ref{sec:rot:thermo}. The particular cases $\nu = 1/2$, $1/3$ and $2/3$ are explicitly discussed in Sec.~\ref{sec:rot:particular}. The massless limit is taken in Sec.~\ref{sec:rot:massless}, establishing agreement with the results in Ref.~\cite{Patuleanu:2025zbn}.

\subsection{Arbitrary rotation parameter} \label{sec:rot:arbitrary}

For states under imaginary rotation, $\widetilde{\mathcal{E}}_\varsigma = \mathcal{E}_\varsigma - i \Omega_I m$ and the Fermi-Dirac factor can be expanded as
\begin{multline}
 \frac{1}{e^{\beta \widetilde{\mathcal{E}}_\varsigma} + 1} = \theta(\mathcal{E}_\varsigma) \sum_{v =1}^\infty (-1)^{v+1} e^{-v \beta \mathcal{E}_\varsigma} e^{i v \beta \Omega_I m} \\
 - \theta(-\mathcal{E}_\varsigma) \sum_{v = 0}^\infty (-1)^{v+1} e^{v \beta \mathcal{E}_\varsigma} e^{-i v \beta \Omega_I m}.
\end{multline}

The $v = 0$ contribution on the second line is temperature- and rotation-independent, corresponding to the degenerate contribution due to the Fermi level set by the chemical potential $\mu$. Writing 
\begin{equation}
 \langle \bar{\psi} \psi \rangle = \langle \bar{\psi} \psi \rangle_{\rm deg} + \Delta \langle \bar{\psi} \psi \rangle, \quad 
 \phi_\psi = \phi_{\rm deg} + \Delta \phi,
\end{equation}
we evaluate the degenerate contributions as follows:
\begin{align}
 \langle \bar{\psi} \psi \rangle_{\rm deg} &= \frac{M N_c N_f}{\pi^2} \sum_{\varsigma = \pm 1} \int_M^\infty dE\, p\, \theta(\varsigma \mu - E) \nonumber\\
 \phi_{\rm deg} &= -\frac{N_c N_f}{\pi^2} \sum_{\varsigma = \pm 1} \int_M^\infty dE\, \frac{p^3}{3}\, \theta(\varsigma \mu - E).
 \label{eq:deg}
\end{align}
In the above, we employed the following summation formula to eliminate the Bessel functions:
\begin{equation}
 \sum_{m = -\infty}^\infty J_m^+(q\rho) = 2.
\end{equation}
The integrals in Eq.~\eqref{eq:deg} can be performed analytically:
\begin{align}
 \langle \bar{\psi} \psi \rangle_{\rm deg} &= \frac{M N_c N_f}{2\pi^2} \left(|\mu| p_f + M^2 \ln \frac{M}{|\mu| + p_f}\right) \theta_M,\nonumber\\
 \phi_{\rm deg} &= -\frac{N_c N_f}{24\pi^2} \bigg[|\mu| p_f(2\mu^2 - 5M^2) \nonumber\\
 & - 3 M^4 \ln \frac{M}{|\mu|+ p_f}\bigg] \theta_M,
 \label{eq:deg_analytic}
\end{align}
with $\theta_M = \theta(|\mu| - M)$ and $p_f = \sqrt{\mu^2 - M^2}$ being the Fermi momentum. The results in Eq.~\eqref{eq:deg_analytic} agree with those in Eq.~(62) of Ref.~\cite{Singha:2025zvh}.

For the non-degenerate contributions, we employ the following summation formula \cite{Patuleanu:2025zbn,Ambrus:2019ayb,Ambrus:2014uqa,DLMF}:
\begin{equation}
 \sum_{m = -\infty}^\infty J_m^+(q\rho) e^{\pm i v \beta \Omega_I m} = 2 c_v J_0(2 q \rho s_v),
 \label{eq:summ_Omega}
\end{equation}
where we introduced the notation
\begin{equation}
 s_v = \sin \left(\frac{v \beta \Omega_I}{2}\right), \quad 
 c_v = \cos \left(\frac{v \beta \Omega_I}{2}\right).
 \label{eq:sv,cv}
\end{equation}

With the aid of the summation formula \eqref{eq:summ_Omega}, we find
\begin{multline}
 \begin{pmatrix}
     \Delta \langle \bar{\psi} \psi \rangle \\
     -\Delta \phi
 \end{pmatrix}
 = \frac{N_c N_f}{2\pi^2} \sum_{\varsigma = \pm 1} \int_M^\infty dE\, {\rm sgn}(\mathcal{E}_\varsigma) \\\times 
 \sum_{v = 1}^\infty (-1)^{v+1} c_v e^{-v \beta |\mathcal{E}_\varsigma|}
 \int_{-p}^p dk 
 \begin{pmatrix}
  M \\ k^2
 \end{pmatrix} J_0(2q \rho s_v).
\end{multline}

The integration with respect to $k$ can be performed using Eqs.~(126)--(129) of Ref.~\cite{Patuleanu:2025zbn}:
\begin{align}
 \int_{-p}^p dk\, J_0(2q \rho s_v) &= \frac{\sin(2p \rho s_v)}{\rho s_v}, \nonumber\\
 \int_{-p}^p dk\, k^2 J_0(2q \rho s_v) &= -\frac{1}{4\rho s_v^2} \frac{d}{d\rho} \left[\frac{\sin(2 p \rho s_v)}{\rho s_v}\right].
\end{align}
This leads to the following result:
\begin{multline}
 \begin{pmatrix}
     \Delta \langle \bar{\psi} \psi \rangle \\
     \Delta \phi 
 \end{pmatrix} = \frac{N_c N_f}{2\pi^2} \sum_{v = 1}^\infty (-1)^{v+1} c_v \begin{pmatrix} 
  M \\ {\displaystyle \frac{1}{4\rho s_v^2} \frac{d}{d\rho}}
 \end{pmatrix} \\\times 
 \sum_{\varsigma = \pm 1} \int_M^\infty dE\, {\rm sgn}(\mathcal{E}_\varsigma) e^{-v\beta |\mathcal{E}_\varsigma|} \frac{\sin(2p \rho s_v)}{\rho s_v}.
 \label{eq:sumv}
\end{multline}

In what follows, it is convenient to write $s_v = \sin(v \pi \nu)$ and $c_v = \cos(v \pi \nu)$, where the dimensionless rotation parameter reads
\begin{equation}
 \nu = \frac{\beta \Omega_I}{2\pi}.
\end{equation}

When $\nu$ is irrational, the trigonometric functions $s_v$ and $c_v$ are non-periodic with respect to $v = 1, 2, \dots$ and Eq.~\eqref{eq:sumv} must be evaluated numerically. In the case when $\nu$ is a rational number, the periodicity of $s_v$ and $c_v$ can be exploited to reveal the emergence of fractal thermodynamics, discussed in the following subsection.

\subsection{Rational rotation parameter} \label{sec:rot:rational}

We now consider the case when the rotation parameter $\nu$ is a rational number, i.e.
\begin{equation}
 \nu = \frac{\beta \Omega_I}{2\pi} = \frac{\mathsf{p}}{\mathsf{q}}, \quad \mathsf{p}, \mathsf{q} \in \mathbb{N}_+, \quad
 {\rm gcd}(\mathsf{p}, \mathsf{q}) = 1,
\end{equation}
where ${\rm gcd}(\mathsf{p},\mathsf{q})$ denotes the greatest common divisor of the integers $\mathsf{p}$ and $\mathsf{q}$ (we take $\mathsf{q} > 0$), i.e.~$\mathsf{p}$ and $\mathsf{q}$ are coprime numbers. Decomposing the summation index $v$ in Eq.~\eqref{eq:sumv} as 
\begin{equation}
 v = \mathsf{q} Q + \mathsf{r},
\end{equation}
with $1 \le \mathsf{r} \le \mathsf{q}$ and $Q = 0, 1, 2, \dots$, 
the trigonometric functions $s_v$ and $c_v$ in Eq.~\eqref{eq:sv,cv} can be simplified to
\begin{align}
 s_v &= (-1)^{\mathsf{p}Q} s_{\mathsf{r}}, & s_{\mathsf{r}} &= \sin\left(\frac{\pi \mathsf{r} \mathsf{p}}{\mathsf{q}}\right), \nonumber\\
 c_v &= (-1)^{\mathsf{p}Q} c_{\mathsf{r}}, & c_{\mathsf{r}} &= \cos\left(\frac{\pi \mathsf{r} \mathsf{p}}{\mathsf{q}}\right).
 \label{eq:sr,cr}
\end{align}

With these properties, the sum over $v$ in  Eq.~\eqref{eq:sumv} can be split via $\sum_{v = 1}^\infty f_v = \sum_{\mathsf{r} = 1}^\mathsf{q} \sum_{Q = 0}^\infty f_{\mathsf{r} + \mathsf{q} Q}$. Taking advantage of the fact that both $s_{\mathsf{r}}$ and $c_{\mathsf{r}}$ are independent of $Q$, we find
\begin{multline}
 \begin{pmatrix}
     \Delta \langle \bar{\psi} \psi \rangle \\
     \Delta \phi 
 \end{pmatrix} = \frac{N_c N_f}{2\pi^2} \sum_{\mathsf{r} = 1}^{\mathsf{q}} (-1)^{\mathsf{r}+1} c_{\mathsf{r}} \begin{pmatrix} 
  M \\ {\displaystyle \frac{1}{4\rho s_{\mathsf{r}}^2} \frac{d}{d\rho}}
 \end{pmatrix} \\\times 
 \sum_{\varsigma = \pm 1} \int_M^\infty dE\, {\rm sgn}(\mathcal{E}_\varsigma) e^{- \mathsf{r} \beta |\mathcal{E}_\varsigma|} \frac{\sin(2 p \rho s_{\mathsf{r}})}{\rho s_{\mathsf{r}}} \\\times 
 \sum_{Q = 0}^\infty (-1)^{\mathsf{k} Q} e^{-\mathsf{q} Q \beta |\mathcal{E}_\varsigma|},
 \label{eq:sumr_aux}
\end{multline}
where we introduced the notation $\mathsf{k} = \mathsf{p} + \mathsf{q}$. The series over $Q$ can be performed analytically via
\begin{equation}
 \sum_{Q = 0}^\infty (-1)^{\mathsf{k} Q} e^{-Q \beta_{\mathsf{q}} |\mathcal{E}_\varsigma|} = \frac{e^{\beta_{\mathsf{q}} |\mathcal{E}_\varsigma|}}{e^{\beta_{\mathsf{q}} |\mathcal{E}_\varsigma|} + (-1)^{\mathsf{k}+1}},
 \label{eq:sumQ}
\end{equation}
where we introduced the fractal inverse temperature 
\begin{equation}
 \beta_{\mathsf{q}} = \mathsf{q} \beta = \frac{1}{T_{\mathsf{q}}}, \quad 
 T_{\mathsf{q}} = \frac{T}{\mathsf{q}},
 \label{eq:Tq}
\end{equation}
and the divergence of the series for $\mathcal{E}_\varsigma = 0$, or $E = \varsigma \mu$, is avoided by taking the Cauchy principal value of the energy integral, as discussed in Appendix~\ref{app:Cauchy}.

The distribution function emerging in Eq.~\eqref{eq:sumQ} corresponds to a fermion (odd $\mathsf{k}$) or boson (even $\mathsf{k}$) gas at an effective temperature $T_\mathsf{q} = T/\mathsf{q}$. Note that the chemical potential $\mu$ remains unaffected by $\mathsf{q}$. 

A remarkable feature, highlighted also in Refs.~\cite{Ambrus:2023bid,Patuleanu:2025zbn}, is that the terms in Eq.~\eqref{eq:sumr_aux} corresponding to $\mathsf{r} = \mathsf{q}$ are independent of the radial coordinate $\rho$. This can be seen since the trigonometric functions $s_{\mathsf{r}}$ and $c_{\mathsf{r}}$ evaluate to
\begin{equation}
 s_{\mathsf{r} = \mathsf{q}} = 0, \quad c_{\mathsf{r} = \mathsf{q}} = (-1)^{\mathsf{p}}.
\end{equation}

We therefore group the degenerate contributions $\langle \bar{\psi}\psi \rangle_{\rm deg}$ and $\phi_{\rm deg}$, which are also coordinate-independent, together with the $\mathsf{r} = \mathsf{q}$ term, by employing the following split:
\begin{equation}
 \langle \bar{\psi}\psi \rangle = \langle \bar{\psi}\psi \rangle_{\mathsf{q}} + \delta \langle \bar{\psi}\psi \rangle, \quad 
 \phi_\psi = \phi_{\mathsf{q}} + \delta \phi.
 \label{eq:fractal_decomposition}
\end{equation}

The coordinate-independent (fractal) contributions are obtained by adding the degenerate contributions from Eq.~\eqref{eq:deg} and the $\mathsf{r} = \mathsf{q}$ contribution from Eq.~\eqref{eq:sumr_aux}:
\begin{multline}
 \begin{pmatrix}
  \langle \bar{\psi}\psi \rangle_{\mathsf{q}}\\
  -\phi_{\mathsf{q}} 
 \end{pmatrix} = \frac{N_c N_f}{\pi^2} \sum_{\varsigma = \pm 1} \int_M^\infty dE\, 
 \begin{pmatrix}
  M p \\ p^3 / 3
 \end{pmatrix} \\
 \times \left[\theta(-\mathcal{E}_\varsigma) + 
 \frac{(-1)^{\mathsf{k}+1} {\rm sgn}(\mathcal{E}_\varsigma)}{e^{\beta_{\mathsf{q}} |\mathcal{E}_\varsigma|} + (-1)^{\mathsf{k}+1}}
 \right].
\end{multline}

Writing ${\rm sgn}(\mathcal{E}_\varsigma) = \theta(\mathcal{E}_\varsigma) - \theta(-\mathcal{E}_\varsigma)$, the degenerate and $\mathsf{r} = \mathsf{q}$ contributions can be combined, yielding
\begin{align}
 \langle \bar{\psi}\psi \rangle_{\mathsf{q}} &= 
 (-1)^{\mathsf{k}+1} \frac{M N_c N_f}{\pi^2} \sum_{\varsigma = \pm 1} \int_M^\infty \frac{dE\, p}{e^{\beta_{\mathsf{q}} \mathcal{E}_\varsigma} + (-1)^{\mathsf{k}+1}}, \nonumber\\
 \phi_{\mathsf{q}} &= (-1)^{\mathsf{k}} \frac{N_c N_f}{3\pi^2} \sum_{\varsigma = \pm 1} \int_M^\infty \frac{dE\, p^3}{e^{\beta_{\mathsf{q}} \mathcal{E}_\varsigma} + (-1)^{\mathsf{k}+1}}.
 \label{eq:fractal_terms}
\end{align}

The transient terms $\delta \langle \bar{\psi} \psi \rangle$ and $\delta \phi$ read
\begin{multline}
 \begin{pmatrix}
     \delta \langle \bar{\psi} \psi \rangle \\
     \delta \phi 
 \end{pmatrix} = \frac{N_c N_f}{2\pi^2} \sum_{\mathsf{r} = 1}^{\mathsf{q} - 1} (-1)^{\mathsf{r}+1} c_{\mathsf{r}} \begin{pmatrix} 
  M \\ {\displaystyle \frac{1}{4\rho s_{\mathsf{r}}^2} \frac{d}{d\rho}}
 \end{pmatrix} \\\times 
 \sum_{\varsigma = \pm 1} \int_M^\infty \frac{dE\, {\rm sgn}(\mathcal{E}_\varsigma) e^{\beta(\mathsf{q} - \mathsf{r}) |\mathcal{E}_\varsigma|}}{e^{\beta_{\mathsf{q}} |\mathcal{E}_\varsigma|} + (-1)^{\mathsf{k}+1}} \frac{\sin(2p \rho s_{\mathsf{r}})}{\rho s_{\mathsf{r}}}.
 \label{eq:sumr}
\end{multline}

It can be seen that the transient terms shown above vanish as $\rho \to \infty$. Thus, in the far-field limit, the expectation values of the scalar condensate and of the thermodynamic potential are dominated by the fractal contributions:
\begin{equation}
 \langle \bar{\psi} \psi \rangle \xrightarrow[\rho \to \infty]{} \langle \bar{\psi} \psi \rangle_\mathsf{q}, \quad 
 \phi_\psi \xrightarrow[\rho \to \infty]{} \phi_\mathsf{q}.
\end{equation}

Before ending this subsection, we remark that the summation over $Q$ leads to a distribution function with statistics that depend on the value of $\mathsf{k} = \mathsf{p} + \mathsf{q}$. This is especially obvious in the far-field limit, where the fractal terms in Eq.~\eqref{eq:fractal_terms} dominate. When $\mathsf{k}$ is odd, the usual Fermi-Dirac statistics is recovered. When $\mathsf{k}$ is even, the distribution is of Bose-Einstein type, being multiplied by an overall factor of $(-1)^{\mathsf{k}+1} = -1$, hence the fermion fields behave like scalar ghosts. At finite quark chemical potential $|\mu| > M$, the phenomenon of Bose-Einstein condensation appears and the distribution develops a (Cauchy-integrable) singularity around $\mathcal{E}_\varsigma = 0$, or $E = |\mu|$. Details can be found in Appendix~\ref{app:Cauchy}.

\subsection{Thermodynamic quantities} \label{sec:rot:thermo}

Starting from the Gibbs-Duhem relation,
\begin{equation}
 d\phi = - s dT - Q d\mu - \mathcal{M}_I d \Omega_I,
\end{equation}
we introduce the entropy density $s$, charge $Q$ and (imaginary) angular momentum $\mathcal{M}_I$ via
\begin{equation}
 s = -\frac{\partial \phi}{\partial T}, \quad 
 Q = -\frac{\partial \phi}{\partial \mu}, \quad 
 \mathcal{M}_I = -\frac{\partial \phi}{\partial \Omega_I}.
\end{equation}
We also introduce at this point the moment of inertia density $\mathcal{I}$, defined for a system under real rotation as
\begin{equation}
 \mathcal{I} = \frac{\mathcal{M}}{\Omega} = -\frac{1}{\Omega} \frac{\partial \phi}{\partial \Omega}.
\end{equation}
In the case of imaginary rotation, $\Omega = i \Omega_I$, we have
\begin{equation}
 \mathcal{I} = \frac{1}{\Omega_I} \frac{\partial \phi}{\partial \Omega_I} = -\frac{\mathcal{M}_I}{\Omega_I}.
 \label{eq:momI_def}
\end{equation}

Since the meson part $V_\sigma$ is independent of the thermodynamic parameters, only the quark part $\phi_\psi$ contributes to the above expressions. Since $\phi_{\rm deg}$ is independent of $T$ and $\Omega_I$, only the charge density receives a degenerate contribution:
\begin{align}
 Q_{\rm deg} &= -\frac{\partial \phi_{\rm deg}}{\partial \mu} = \frac{N_c N_f}{\pi^2} \sum_{\varsigma = \pm 1} \varsigma \int_M^\infty dE\, p E \theta(-\mathcal{E}_\varsigma) \nonumber\\
 &= \frac{N_c N_f p_f^3}{3\pi^2} {\rm sgn}(\mu),
 \label{eq:deg_Q}
\end{align}
where we considered the relation $\partial f(\mathcal{E}_\varsigma) /\partial \mu = -\varsigma \partial f(\mathcal{E}_\varsigma) / \partial E$, followed by an integration by parts in Eq.~\eqref{eq:deg}.

The nondegenerate contributions can be obtained by differentiating $\Delta \phi$ in Eq.~\eqref{eq:sumv}:
\begin{align}
 \Delta s &= -\frac{N_c N_f}{2\pi^2} \sum_{v = 1}^\infty \frac{(-1)^{v+1}}{4\rho} \frac{d}{d\rho} \frac{1}{\rho} \sum_{\varsigma = \pm 1} \int_M^\infty dE \nonumber\\
 &\times {\rm sgn}(\mathcal{E}_\varsigma) \frac{d}{dT} \left[e^{-v\beta |\mathcal{E}_\varsigma|} \frac{c_v}{s_v^3} \sin(2p \rho s_v) \right], \nonumber\\
 \Delta Q &= -\frac{N_c N_f}{2\pi^2} \sum_{v = 1}^\infty \frac{(-1)^{v+1}}{4\rho} \frac{d}{d\rho} \frac{1}{\rho} \sum_{\varsigma = \pm 1} \varsigma \int_M^\infty dE \, \nonumber\\
 &\times {\rm sgn}(\mathcal{E}_\varsigma) e^{-v\beta |\mathcal{E}_\varsigma|} \frac{c_v}{s_v^3} \frac{d}{dE} \sin(2p \rho s_v), \nonumber\\
 \Delta \mathcal{M}_I &= \frac{N_c N_f T}{2\pi^2 \Omega_I} \sum_{v = 1}^\infty \frac{(-1)^{v+1}}{4\rho} \frac{d}{d\rho} \frac{1}{\rho} \sum_{\varsigma = \pm 1} \int_M^\infty dE \nonumber\\
 &\times {\rm sgn}(\mathcal{E}_\varsigma) e^{-v\beta |\mathcal{E}_\varsigma|} \frac{d}{dT} \left[\frac{c_v}{s_v^3} \sin(2p \rho s_v)\right],
\label{eq:thermodynamic_vsum}
\end{align}
where we used the property $\partial f(\beta \Omega_I) / \partial \Omega_I = -\frac{T}{\Omega_I} \partial f / \partial T$ and we took into account that the entire dependence on $\Omega_I$ is contained in the terms $s_v$ and $c_v$ defined in Eq.~\eqref{eq:sv,cv}.

Using the Euler relation,
\begin{equation}
 s T = \epsilon - \phi - \mu Q - \Omega_I \mathcal{M}_I,
\end{equation}
we can obtain the expression for the energy density $\epsilon = \epsilon_{\rm deg} + \Delta \epsilon$. The degenerate contribution can be obtained by multiplying Eq.~\eqref{eq:deg_Q} by $\mu$ and adding $\phi_{\rm deg}$ given in Eq.~\eqref{eq:deg}:
\begin{align}
 \epsilon_{\rm deg} &= \frac{N_c N_f}{\pi^2} \sum_{\varsigma =\pm 1}  \int_M^\infty dE\, pE^2 \theta(-\mathcal{E}_\varsigma) \nonumber\\
 &= \frac{N_c N_f}{8\pi^2} \left[|\mu| p_f (2\mu^2 - M^2) + M^4 \ln \frac{M}{|\mu| + p_f}\right].
\end{align}

The nondegenerate contribution can be obtained as:
\begin{align}
 \Delta \epsilon &= T \Delta s + \Delta \phi + \mu \Delta Q + \Omega_I \Delta \mathcal{M}_I \nonumber\\
 &= \frac{N_c N_f}{2\pi^2} \sum_{v = 1}^\infty \frac{(-1)^{v+1} c_v}{4\rho s_v^3} \frac{d}{d\rho} \frac{1}{\rho} \sum_{\varsigma = \pm 1} \int_M^\infty dE\, {\rm sgn}(\mathcal{E}_\varsigma) \nonumber\\
 &\times \left[\sin(2p \rho s_v) \left(e^{-v\beta |\mathcal{E}_\varsigma|} - T \frac{d}{dT} e^{-v\beta |\mathcal{E}_\varsigma|}\right) \right. \nonumber\\
 & \left. - (E - \mathcal{E}_\varsigma) e^{-v \beta |\mathcal{E}_\varsigma|} \frac{d}{dE} \sin(2p \rho s_v)\right],
\end{align}
where we wrote $\varsigma \mu = E - \mathcal{E}_\varsigma$.
Noting that $-T d e^{-v \beta |\mathcal{E}_\varsigma|} / dT = \mathcal{E}_\varsigma de^{-v \beta |\mathcal{E}_\varsigma|} / dE$ and inserting $1 = d\mathcal{E}_\varsigma / dE$ in the first term, the first, second and fourth terms form the differential $d[\mathcal{E}_\varsigma e^{-v\beta|\mathcal{E}_\varsigma|} \sin(2p \rho s_v)] / dE$, which makes a vanishing contribution under the integration with respect to $E$. We are thus left with 
\begin{multline}
 \Delta \epsilon = \frac{N_c N_f}{2\pi^2} \sum_{v = 1}^\infty (-1)^{v+1} c_v \sum_{\varsigma = \pm 1} \int_M^\infty dE \, E^2\, {\rm sgn}(\mathcal{E}_\varsigma) \\\times 
 e^{-v\beta |\mathcal{E}_\varsigma|} \frac{\sin(2p \rho s_v)}{\rho s_v}.
\end{multline}

\subsubsection*{Rational rotation frequency}

We now consider the case of rational rotation parameter, $\nu = \mathsf{p} / \mathsf{q}$. Following the same steps as described in Subsec.~\ref{sec:rot:rational}, we obtain the  transient terms as follows:
\begin{multline}
 \begin{pmatrix}
  \delta \epsilon \\ 
  \delta Q \\ 
  \delta \mathcal{M}_I
 \end{pmatrix} = \frac{N_c N_f}{2\pi^2} \sum_{\mathsf{r} = 1}^{\mathsf{q} - 1} (-1)^{\mathsf{r}+1} \sum_{\varsigma = \pm 1} \int_M^\infty dE\, E \\\times 
 \frac{e^{(\mathsf{q} - \mathsf{r}) \beta |\mathcal{E}_\varsigma|}}{e^{\beta_{\mathsf{q}} |\mathcal{E}_\varsigma|} + (-1)^{\mathsf{k}+1}}
 \begin{pmatrix}
     E {\rm sgn}(\mathcal{E}_\varsigma) \\ \varsigma {\rm sgn}(\mathcal{E}_\varsigma) \\ 
     {\displaystyle \frac{c_{\mathsf{r}}}{2} \frac{d}{ds_{\mathsf{r}}}} 
 \end{pmatrix} c_{\mathsf{r}} \frac{\sin(2 p \rho s_{\mathsf{r}})}{\rho s_{\mathsf{r}}},
 \label{eq:sumr_thermo}
\end{multline}
while $\delta s$ can be obtained either via the Euler relation, $\delta s = (\delta \epsilon - \delta \phi - \mu \delta Q - \Omega_I \delta \mathcal{M}_I) / T$, or by $\delta s = -d \delta \phi / dT$, where one must take into account that the temperature derivative acts also on the functions $c_{\mathsf{r}}$ and $s_{\mathsf{r}}$ in Eq.~\eqref{eq:sumr}.

For even $\mathsf{k}$, the evaluation of the angular-momentum contribution in Eq.~\eqref{eq:sumr_thermo} requires additional care when $|\mu|>M$.
In this case, the ghost-like distribution becomes singular at $\mathcal{E}_\varsigma=0$, and the individual terms in the sum over $\mathsf{r}$ contributing to $\delta\mathcal{M}_I$ are not separately integrable. In contrast to the other thermodynamic quantities in Eq.~\eqref{eq:sumr_thermo}, whose singular energy integrals can be defined in the Cauchy principal-value sense, as in the case of free energy discussed in Appendix~\ref{app:Cauchy}, the Cauchy principal value of an individual $\mathsf{r}$ contribution to $\delta\mathcal{M}_I$ does not exist. The singularities cancel only when the corresponding terms in the sum over $\mathsf{r}$ are combined. Following the procedure of Ref.~\cite{Patuleanu:2025zbn} [see Eq.~(B10) therein], this cancellation can be made explicit by pairing the contributions with indices $\mathsf{r}$ and $\mathsf{q}-\mathsf{r}$ before performing the energy integration.
The resulting paired integrand in the expression of $\delta \mathcal{M}_I$ is regular at $\mathcal{E}_\varsigma=0$.

The non-transient (fractal) contributions can be obtained as follows:
\begin{multline}
 \begin{pmatrix}
  \epsilon_\mathsf{q} \\ Q_\mathsf{q} \\ s_\mathsf{q} T
 \end{pmatrix} =\frac{N_c N_f}{\pi^2} \sum_{\varsigma = \pm 1} \int_M^\infty dE \begin{pmatrix}
    pE^2 \\ \varsigma p E \\ 
    \frac{1}{3} p^3 + pE \mathcal{E}_{\varsigma}\smallskip 
 \end{pmatrix} \\\times 
  \frac{(-1)^{\mathsf{k}+1}}{e^{\beta_\mathsf{q} \mathcal{E}_\varsigma} + (-1)^{\mathsf{k}+1}},
\end{multline}
while $\mathcal{M}_{I;\mathsf{q}} = 0$. It can be readily checked that 
\begin{equation}
 s_{\mathsf{q}} = -\frac{\partial \phi_{\mathsf{q}}}{\partial T}.
 \label{eq:fractal_terms_s}
\end{equation}

While this relation seems straightforward from a thermodynamic point of view, it reveals a subtle inconsistency in the far-field thermodynamics. Let us consider for the moment the case when $\mathsf{k}$ is odd. The expressions for $\phi_{\mathsf{q}}$ \eqref{eq:fractal_terms}, $\epsilon_{\mathsf{q}}$ and $Q_{\mathsf{q}}$ coincide with those for the free energy, energy and charge densities of a Fermi gas at rest, at temperature $T_{\mathsf{q}}$ and chemical potential $\mu$. However, Eq.~\eqref{eq:fractal_terms_s} reveals an entropy density which is $T / T_{\mathsf{q}} = \mathsf{q}$ times smaller than that imposed by thermal equilibrium, $s_{\mathsf{q}}^{eq} = -\partial \phi_{\mathsf{q}} / \partial T_{\mathsf{q}} = \mathsf{q} s_{\mathsf{q}}$. Thus, we can conclude that the system does not reach a consistent thermal equilibrium state in the far-field limit.

\subsection{Particular cases} \label{sec:rot:particular}

The discussions in Subsecs.~\ref{sec:rot:rational} and \ref{sec:rot:thermo} reveal that the thermal expectation values can be written as a sum of a fractal, coordinate-independent contribution and $\mathsf{q} - 1$ transient terms. In this subsection, we evaluate the transient terms for a few special cases.

When $\nu = 1/2$ ($\mathsf{p} = 1$ and $\mathsf{q} = 2$), the summation index $\mathsf{r}$ takes only the value $\mathsf{r} = 1$, in which case the trigonometric functions $s_{\mathsf{r}}$ and $c_{\mathsf{r}}$ in Eq.~\eqref{eq:sr,cr} evaluate to
\begin{equation}
 s_{\mathsf{r} = 1} = 1, \quad c_{\mathsf{r} = 1} = 0, \qquad 
 (\mathsf{p}, \mathsf{q}) = (1,2).
\end{equation}
In this case, all transient quantities in Eqs.~\eqref{eq:sumr} and \eqref{eq:sumr_thermo} vanish due to the leading $c_{\mathsf{r}}$ factor inside the sum over $\mathsf{r}$, except the angular momentum density, which evaluates to
\begin{equation}
 \mathcal{M}_I\rvert_{\substack{\mathsf{p} = 1\\\mathsf{q} = 2}} = -\frac{N_c N_f}{8\pi^2 \rho} \sum_{\varsigma = \pm 1} \int_M^\infty \frac{dE\, E}{\cosh(\beta \mathcal{E}_\varsigma)} \sin(2 p \rho).
 \label{eq:MI_1_over_2}
\end{equation}

In the case $(\mathsf{p}, \mathsf{q}) = (1,3)$ we have an even $\mathsf{k} = \mathsf{p} + \mathsf{q}$. We find:
\begin{align}
 \begin{pmatrix}
     \delta \langle \bar{\psi} \psi \rangle  \\ \delta Q \\ \delta \epsilon \\ \delta \phi
 \end{pmatrix}
 &= \frac{N_c N_f}{2\pi^2} \sum_{\varsigma = \pm 1} \int_M^\infty \frac{dE\,\cosh(\frac{\beta}{2} \mathcal{E}_\varsigma)}{\sinh(\frac{3}{2}\beta \mathcal{E}_\varsigma)} 
 \begin{pmatrix}
     M \\ \varsigma E \\ E^2 \\ \frac{1}{3\rho} \frac{d}{d\rho}
 \end{pmatrix} \nonumber\\
 & \times \frac{\sin(p \rho \sqrt{3})}{\rho \sqrt{3}}, \nonumber\\
 \delta \mathcal{M}_I &= -\frac{N_c N_f}{2\pi^2} \sum_{\varsigma = \pm 1} \int_M^\infty \frac{dE\, E\, \sinh(\frac{\beta}{2} \mathcal{E}_\varsigma)}{\sinh(\frac{3}{2} \beta \mathcal{E}_\varsigma)} \nonumber\\
 & \times \left[\frac{2}{3\rho} \sin(p \rho \sqrt{3}) - \frac{p}{2\sqrt{3}} \cos(p \rho \sqrt{3})\right].
\end{align}

When $(\mathsf{p}, \mathsf{q}) = (2,3)$, we get
\begin{align}
 \begin{pmatrix}
   \delta \langle \bar{\psi} \psi \rangle  \\ \delta Q \\ \delta \epsilon \\ \delta \phi
 \end{pmatrix}
 & = -\frac{N_c N_f}{2\pi^2} \sum_{\varsigma = \pm 1} \int_M^\infty \frac{dE\, \sinh(\frac{\beta}{2} \mathcal{E}_\varsigma)}{\cosh(\frac{3}{2}\beta \mathcal{E}_\varsigma)} \begin{pmatrix}
     M \\ \varsigma E \\ E^2 \\ \frac{1}{3\rho} \frac{d}{d\rho}
 \end{pmatrix} \nonumber\\ 
 & \times \frac{\sin(p \rho \sqrt{3})}{\rho \sqrt{3}}, \nonumber\\
 \delta \mathcal{M}_I &= -\frac{N_c N_f}{2\pi^2} \sum_{\varsigma = \pm 1} \int_M^\infty \frac{dE\, E \cosh(\frac{1}{2} \beta \mathcal{E}_\varsigma)}{\cosh(\frac{3}{2} \beta \mathcal{E}_\varsigma)} \nonumber\\
 & \times \left[\frac{2}{3\rho} \sin(p \rho \sqrt{3}) - \frac{p}{2\sqrt{3}} \cos(p \rho \sqrt{3})\right],
\end{align}

\subsection{Massless limit} \label{sec:rot:massless}

In this subsection, we validate our results obtained so far with those derived in Ref.~\cite{Patuleanu:2025zbn} for the case of a single, massless fermion species.
In the massless limit $M \to 0$, the degenerate contributions from Eq.~\eqref{eq:deg_analytic} simplify to:
\begin{align}
 \langle \bar{\psi} \psi \rangle_{\rm deg} &\simeq \frac{M N_c N_f}{2\pi^2} \mu^2, &
 \phi_{\rm deg} &\simeq -\frac{N_c N_f}{12\pi^2} \mu^4,
 \label{eq:massless_deg}
\end{align}
in agreement with Eqs.~(113) and (191) of Ref.~\cite{Patuleanu:2025zbn}.

To obtain the massless limit of the nondegenerate contributions in
Eq.~\eqref{eq:sumv}, we set
$p\to E$ and express the energy integral as
\begin{multline}
\sum_{\varsigma = \pm 1} \int_0^\infty dE\,
{\rm sgn}(\mathcal E_\varsigma)\,
e^{-v\beta|\mathcal E_\varsigma|}
\sin(2E\rho s_v) \\
= \frac{1}{i v \beta} [I_+(\alpha_v) - I_+(-\alpha_v)],
\label{eq:massless_energy_integrals}
\end{multline}
where $I_+(\alpha_v)$ is defined in Eq.~(135) of Ref.~\cite{Patuleanu:2025zbn} by
\begin{equation}
 I_+(\alpha_v) = \frac{1}{2} \sum_{\varsigma = \pm 1} \int_0^\infty dx\, e^{-|X^\varsigma_v|} {\rm sgn}(X^\varsigma_v) e^{i \alpha_v x},
\end{equation}
where $x$, $X^\varsigma_v$ and $\alpha_v$ are given by
\begin{equation}
    x =v\beta E, \quad 
    X^\varsigma_v = x - v \varsigma \beta \mu, \quad 
    \alpha_v=\frac{2\rho s_v}{v\beta}.    
\end{equation}
Using the result in Eq.~(140a) of Ref.~\cite{Patuleanu:2025zbn}, 
\begin{equation}
 I_+(\alpha_v) = \frac{1}{1 + \alpha_v^2} (i \alpha_v e^{i v\beta |\mu| \alpha_v} + e^{-v \beta |\mu|}),
\end{equation}
we arrive at 
\begin{equation}
 \sum_{\varsigma = \pm 1} \int_0^\infty dE\, {\rm sgn}(\mathcal E_\varsigma)\,
e^{-v\beta|\mathcal E_\varsigma|}
\frac{\sin(2E\rho s_v)}{\rho s_v}
=\frac{4}{(v\beta)^2}
\frac{c^{v\mu}}{1+\alpha_v^2},
\end{equation}
where $c^{v\mu}=\cos(2\rho\mu s_v)$.
Thus, the massless nondegenerate contributions to the fermion condensate and thermodynamic potential from Eq.~\eqref{eq:sumv} take the following form:
\begin{equation}
\begin{pmatrix}
\Delta\langle\bar{\psi}\psi\rangle
\\
\Delta\phi
\end{pmatrix}
\simeq
\frac{2N_cN_f}{\pi^2\beta^2}
\sum_{v=1}^\infty
\frac{(-1)^{v+1}}{v^2}
\begin{pmatrix}
M
\\
{\displaystyle
\frac{1}{4\rho s_v^2}\frac{d}{d\rho}}
\end{pmatrix}
\frac{c_v c^{v\mu}}
{1+\alpha_v^2},
\label{eq:massless_sumv}
\end{equation}
in agreement with Eqs.~(144a) and (192) of Ref.~\cite{Patuleanu:2025zbn}.

\subsubsection*{Rational rotation frequency}

We now consider the massless limit for a rational rotation parameter, $\nu = \mathsf{p} / \mathsf{q}$. Applying the same steps as in Sec.~\ref{sec:rot:rational} to the expressions in Eq.~\eqref{eq:massless_sumv}. Writing $v = \mathsf{q} Q + \mathsf{r}$ and selecting only the contributions for $1 \le \mathsf{r} \le \mathsf{q} - 1$ gives the following transient terms:
\begin{multline}
\begin{pmatrix}
\delta\langle\bar{\psi}\psi\rangle
\\[2mm]
\delta\phi
\end{pmatrix}
\simeq
\frac{2N_cN_f}{\pi^2 \beta_\mathsf{q}^2}
\sum_{\mathsf{r}=1}^{\mathsf{q}-1}
(-1)^{\mathsf{r}+1} c_{\mathsf{r}}
\begin{pmatrix}
M
\\[2mm]
{\displaystyle
\frac{1}{4\rho s_{\mathsf{r}}^2}\frac{d}{d\rho}}
\end{pmatrix}
\\\times
\left[
c^{\mathsf{r}\mu}\,
\mathcal Q_\mathsf{k}^\mathsf{r}(x_{\mathsf{r}})
\right],
\label{eq:massless_fractalized}
\end{multline}
where we introduced the following notation:
\begin{gather}
x_\mathsf{r}=\frac{2\rho s_\mathsf{r}}{\mathsf{q}\beta},
\quad
c^{\mathsf{r}\mu}=\cos(2\rho\mu s_\mathsf{r}), \nonumber\\
\mathcal Q_\mathsf{k}^\mathsf{r}(x_\mathsf{r})
=
\sum_{Q=0}^\infty
\frac{(-1)^{\mathsf{k}Q}}
{\left(Q+\frac{\mathsf{r}}{\mathsf{q}}\right)^2 + x_\mathsf{r}^2}.
\end{gather}

The coordinate-independent (fractal) terms are obtained by summing the degenerate contributions in Eq.~\eqref{eq:massless_deg} together with the $\mathsf{r} = \mathsf{q}$ contribution to Eq.~\eqref{eq:massless_sumv}:
\begin{align}
\langle\bar{\psi}\psi\rangle_{\mathsf{q}}
&\simeq 
\frac{M N_cN_f}{\pi^2}
\left[
\frac{\mu^2}{2}
+
2 T_{\mathsf{q}}^2 \mathcal Z_{\mathsf{k}}^{(2)}
\right],\nonumber \\
\phi_{\mathsf{q}}
&\simeq
-\frac{N_cN_f}{\pi^2}
\left[
\frac{\mu^4}{12}
+
2 \mu^2 T_{\mathsf{q}}^2\mathcal Z_{\mathsf{k}}^{(2)}
+
4 T_{\mathsf{q}}^4\mathcal Z_{\mathsf{k}}^{(4)}
\right],
\label{eq:fractal_massless}
\end{align}
where we introduced the notation
\begin{equation}
\mathcal Z_{\mathsf{k}}^{(n)}
\equiv
(-1)^{\mathsf{k}+1} \sum_{Q=0}^{\infty}
\frac{(-1)^{\mathsf{k} Q}}{(Q+1)^n},
\end{equation}
with $\mathcal Z_{\mathsf{k}}^{(2)} = (-1)^{\mathsf{k}+1} \mathcal{Q}^{\mathsf{q}}_{\mathsf{k}} (0)$ and $\mathcal Z_{\mathsf{k}}^{(4)}= \frac{(-1)^{\mathsf{k}}}{2}\left.\frac{d^2\mathcal Q_{\mathsf{k}}^{\mathsf{q}}}{dX^2}\right\vert_{X=0}$, which, depending on the parity of $\mathsf{k} = \mathsf{p} + \mathsf{q}$, evaluate to:
\begin{align}
 \text{odd } \mathsf{k}:& & \mathcal{Z}_{\mathsf{k}}^{(2)} &= \frac{\pi^2}{12}, & 
 \mathcal{Z}_{\mathsf{k}}^{(4)} &= \frac{7\pi^4}{720}, \nonumber\\
 \text{even } \mathsf{k}:& & \mathcal{Z}_{\mathsf{k}}^{(2)} &= -\frac{\pi^2}{6}, & 
 \mathcal{Z}_{\mathsf{k}}^{(4)} &= -\frac{\pi^4}{90}. 
\label{eq:Zk24}
\end{align}
The results obtained in Eqs.~\eqref{eq:massless_fractalized} and \eqref{eq:fractal_massless} above are in agreement with those in Eqs.~(202b), (203b) and (176) of Ref.~\cite{Patuleanu:2025zbn}.

The other thermodynamic quantities ($\epsilon$, $Q$, $\mathcal{M}_I$) can be evaluated in a similar fashion, or by directly differentiating the thermodynamic potential density $\phi_\psi$. In all cases, we find agreement with the results in Ref.~\cite{Patuleanu:2025zbn}.

\section{Phase diagram} \label{sec:pd}

In this section, we discuss the (potentially inhomogeneous) thermodynamic phases in which the system resides, as a function of the temperature $T$, chemical potential $\mu$ and imaginary angular velocity $\Omega = i\Omega_I$ of the grand canonical ensemble. We begin with the discussion of the far-field and rotation axis limits, which form the subject of Subsecs.~\ref{sec:pd:far} and \ref{sec:pd:axis}. We discuss the transition between these two limits as the distance to the rotation axis is varied in Subsec.~\ref{sec:pd:l}.

\subsection{Far-field limit}\label{sec:pd:far}

In the far-field limit $\rho \to \infty$, the thermodynamics of the system is governed by the fractal contributions $\phi_\mathsf{q}$ and $\langle \bar{\psi} \psi \rangle_\mathsf{q}$ to the free energy and fermion condensate, shown in Eq.~\eqref{eq:fractal_terms}. The gap equation \eqref{eq:sigma_eq} becomes
\begin{align}
 \lambda(\sigma_\infty^2 - v^2) \sigma_\infty &= h - g \langle \bar{\psi} \psi \rangle_{\mathsf{q}}.
\end{align}

A first consequence of fractalization is that the effective temperature is reduced by a factor of $\mathsf{q}$, i.e. $T_{\mathsf{q}} = T/ \mathsf{q}$, as discussed in Eq.~\eqref{eq:Tq}. When $\mathsf{k} = \mathsf{p} + \mathsf{q}$ is an odd number, the fermion condensate is positive and it corresponds to that of a static system at temperature $T_\mathsf{q}$ and chemical potential $\mu$. This implies that the transition line $(T_{c; \mathsf{q}}, \mu_{c; \mathsf{q}})$ in the $(T, \mu)$ plane is related to the line $(T_{c;{\rm static}}, \mu_{c;{\rm static}})$ of the non-rotating system by
\begin{equation}
 T_{c; \mathsf{q}} = \mathsf{q} T_{c;{\rm static}}, \quad \mu_{c; \mathsf{q}} = \mu_{c;{\rm static}},
\end{equation}
being obtained from the line corresponding to the static case by a simple rescaling of the temperature.

\begin{figure}
    \centering
    \includegraphics[width=.95\linewidth]{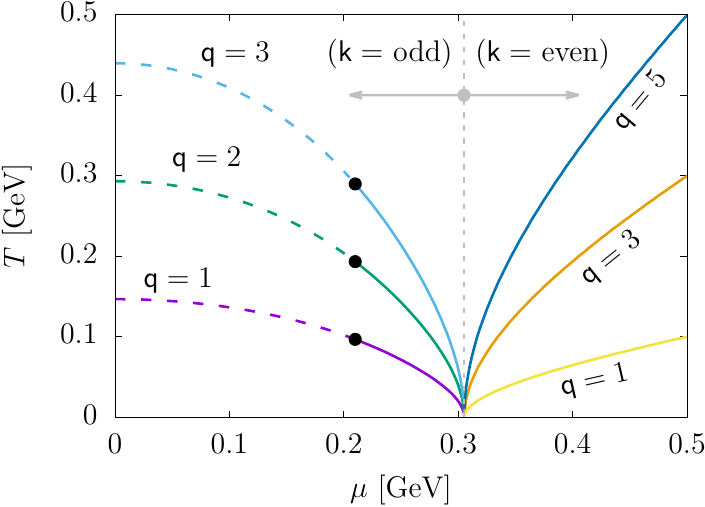}
    \caption{Phase diagram for the system under imaginary rotation in the far-field limit. Each curve corresponds to a given choice of $\mathsf{q}$ and odd (left) or even (right) values of $\mathsf{k} = \mathsf{p} + \mathsf{q}$. The solid (dotted) segments indicate first-order (crossover) phase transitions, while the black circles indicate the critical points corresponding to second-order phase transitions. }
    \label{fig:pd_farfield}
\end{figure}

In the case when $\mathsf{k} = \mathsf{p} + \mathsf{q}$ is even, the sign of the fermion condensate is in principle flipped. On the other hand, Eq.~\eqref{eq:deg} shows that the temperature-independent part of $\langle \bar{\psi} \psi \rangle$ does not suffer from fractalization issues, maintaining a positive sign regardless of the sign of $\mathsf{k}$. Recalling the massless limit for the fermion condensate, derived in Eqs.~(176) of Ref.~\cite{Patuleanu:2025zbn},
\begin{align}
 \text{odd } \mathsf{k}:& & \langle \bar{\psi} \psi \rangle_{\mathsf{q}} &\simeq M N_c N_f \left(\frac{T_{\mathsf{q}}^2}{6} + \frac{\mu^2}{2\pi^2}\right), \nonumber\\
 \text{even } \mathsf{k}:& & \langle \bar{\psi} \psi \rangle_{\mathsf{q}} &\simeq M N_c N_f \left(-\frac{T_{\mathsf{q}}^2}{3} + \frac{\mu^2}{2\pi^2}\right),
\label{eq:FC_far_field_m0}
\end{align}
as well as that for the free energy density (see Eqs.~(202b) and (203b) in Ref.~\cite{Patuleanu:2025zbn}),
\begin{align}
 \text{odd } \mathsf{k}:& & \phi_{\mathsf{q}} &\simeq -N_c N_f \left(\frac{7\pi^2 T_{\mathsf{q}}^4}{180} + \frac{\mu^2 T_{\mathsf{q}}^2}{6} + \frac{\mu^4}{12\pi^2}\right), \nonumber\\
 \text{even } \mathsf{k}:& & \phi_{\mathsf{q}} &\simeq -N_c N_f \left(-\frac{2\pi^2 T_{\mathsf{q}}^4}{45} - \frac{\mu^2 T_{\mathsf{q}}^2}{3} + \frac{\mu^4}{12\pi^2}\right),
\end{align}
we see that the signs of all temperature-dependent contribution are flipped when $\mathsf{k}$ is even. Thus, we are led to believe that, at even $\mathsf{k}$, increasing the temperature of the system inhibits the chiral restoration transition in the far-field limit. The role of the chemical potential remains unchanged compared to the non-rotating case: increasing the chemical potential restores the chiral symmetry. 

The above features are illustrated in Fig.~\ref{fig:pd_farfield}, where the transition line is plotted for various values of $\mathsf{q}$ and for the two cases of odd and even $\mathsf{k}$. For odd $\mathsf{k}$, the closed region in the lower-left side of the plot is in the chirally broken phase. For even $\mathsf{k}$, the closed region in the bottom-right side of the plot is in the chirally restored phase. Since at vanishing temperature, $\langle \bar{\psi} \psi \rangle_{\rm deg}$ \eqref{eq:deg} is independent of $\Omega_I$, all transition lines meet at the point $(T_c, \mu_c) = (0, \mu^c_{T = 0})$, with $\mu^c_{T = 0} \simeq 305\ {\rm MeV}$. When $\mathsf{q} \to \infty$, the transition line approaches the vertical line $\mu_c = 305\ {\rm MeV}$, regardless of the value of $\mathsf{k}$ and thus of $\mathsf{p}$. Therefore, the far-field phase of the system becomes independent of $T$, being chirally broken for $\mu < \mu^c_{T = 0}$ and chirally restored for $\mu > \mu^c_{T = 0}$.

A notable paradox emerges from the above discussion. The absence of imaginary rotation corresponds to $\nu = 0 = 0/1$, with $\mathsf{p} = 0$ and $\mathsf{q} = 1$, as implied by the first relation in Eq.~\eqref{eq:FC_far_field_m0}. The $\mathsf{q} \to \infty$ limit of $\nu = \mathsf{p} / \mathsf{q}$ seemingly also corresponds to the vanishing rotation limit. However, in this case, the effect of the temperature is completely suppressed. 

\subsection{Rotation axis}\label{sec:pd:axis}

\begin{figure}
\centering
\begin{tabular}{c}
\includegraphics[width=0.95\linewidth]{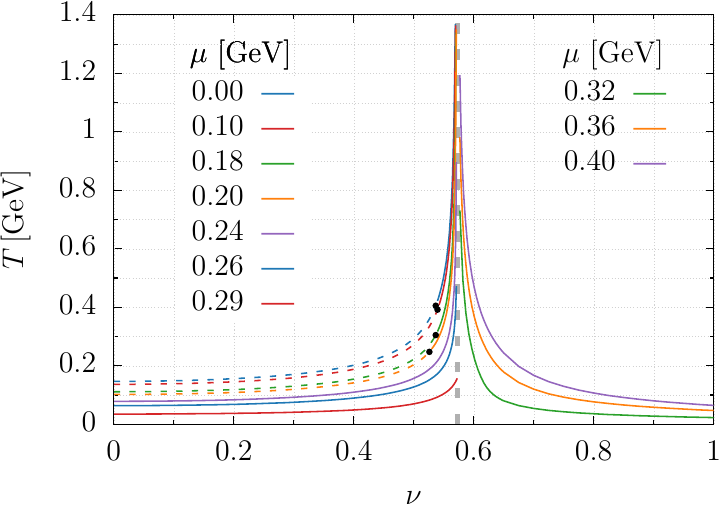} \vspace{-60pt} \\ 
\hspace{-.55\linewidth} (a) \vspace{50pt}\\
\includegraphics[width=0.95\linewidth]{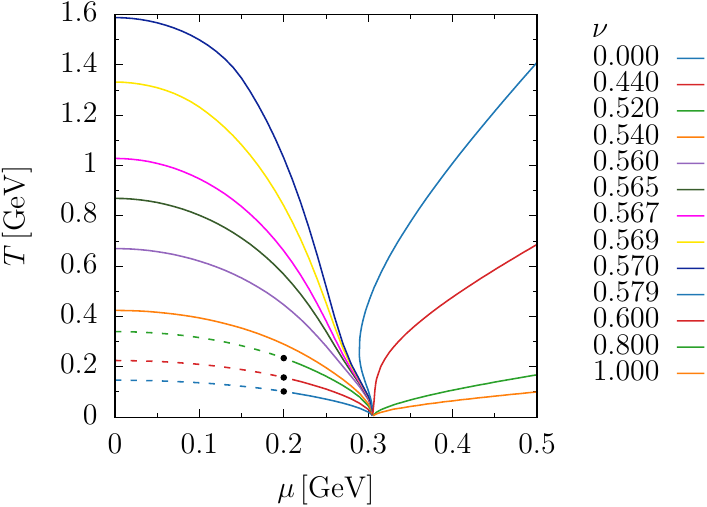} \vspace{-30pt} \\ 
\hspace{.8\linewidth} (b) \vspace{30pt}
\end{tabular}
\caption{Phase diagram on the rotation axis ($\rho = 0$) (a) in the $T$-$\nu$ plane for various values of $\mu$; and (b) in the $T$-$\mu$ plane for various values of $\nu$.}
\label{fig:pd_axis}
\end{figure}

We now discuss the thermodynamic properties of the system on the axis of rotation, when $\rho = 0$. The fermion condensate and grand potential density can be obtained directly from Eqs.~\eqref{eq:SC_gen} and \eqref{eq:phi_gen} by noting that $J_m^+(0) = 1$ for $m = \pm \frac{1}{2}$ and vanishes otherwise:
\begin{multline}
 \begin{pmatrix}
  \langle \bar{\psi} \psi \rangle \\
  -\phi_\psi
 \end{pmatrix}_{\rho = 0}
 = \frac{N_c N_f}{2\pi^2} \sum_{\varsigma = \pm 1} \int_M^\infty dE\, \begin{pmatrix}
     p M \\ p^3 / 3
 \end{pmatrix} \\
 \times \left[\frac{1}{e^{\beta \mathcal{E}_\varsigma - \frac{i}{2} \beta \Omega_I} + 1} + \frac{1}{e^{\beta \mathcal{E}_\varsigma + \frac{i}{2} \beta \Omega_I} + 1}\right].
\end{multline}
The moment of inertia density can be obtained via Eq.~\eqref{eq:momI_def}, by differentiating $\phi_\psi$ given in Eq.~\eqref{eq:phi_ln}:
\begin{align}
 \mathcal{I}\rvert_{\rho = 0} &= -\frac{i \beta N_c N_f}{4 \pi^3 \nu} \sum_{\varsigma = \pm 1} 
 \sum_{m = \pm \frac{1}{2}} m \int_M^\infty \frac{dE\, p E}{e^{\beta \widetilde{\mathcal{E}}_\varsigma} + 1} \nonumber\\
 &= \frac{\beta N_c N_f}{8\pi^2} \frac{\sin(\pi \nu)}{\pi \nu} \sum_{\varsigma = \pm 1} \int_M^\infty \frac{dE\, pE}{\cosh(\beta \mathcal{E}_\varsigma) + \cos(\pi \nu)}.
\end{align}

In the small mass limit, these quantities are given by Eqs.~(149a) and (149e) in Ref.~\cite{Patuleanu:2025zbn}:
\begin{subequations}
\begin{align}
 \langle \bar{\psi}\psi \rangle_{\rho = 0} &\simeq M N_c N_f \left[\frac{T^2}{6} (1 - 3 \nu^2) + \frac{\mu^2}{2\pi^2}\right], \label{eq:axis_ppsi}\\
 \phi_\psi\rvert_{\rho = 0} &\simeq -N_c N_f \left[\frac{7 \pi^2 T^4}{180} \left(1 - \frac{30\nu^2}{7} + \frac{15 \nu^4}{7} \right) \right. \nonumber\\
 & \left. + \frac{\mu^2 T^2}{6} (1 -3\nu^2) + \frac{\mu^4}{12\pi^2}\right], \label{eq:axis_phi}\\
 \mathcal{I}\rvert_{\rho = 0} &\simeq N_c N_f \left[\frac{T^2}{12}(1 - \nu^2) + \frac{\mu^2}{4\pi^2}\right].\label{eq:axis_I}
\end{align}
\end{subequations}
Since the fermion system described by Eq.~\eqref{eq:phi_gen} is periodic with respect to $\nu \to \nu + 2$, with the principal determination given by $-1 \le \nu \le 1$. Within this domain, we see that the moment of inertia density $\mathcal{I}$, corresponding to the spin part identified in Refs.~\cite{Ahadi:2025rqs,Patuleanu:2025zbn}, remains positive for all values of $\nu$.

It can be seen that the coefficient of the thermal term in the fermion condensate depends on $\nu$. Moreover, when $\nu > \nu_c = 1/\sqrt{3} \simeq 0.577$, the sign of the thermal contribution is flipped and therefore increasing the temperature inhibits chiral restoration. 

Figure~\ref{fig:pd_axis} shows the phase diagram on the rotation axis. Panel (a), showing the transition line in the $T$-$\nu$ plane, confirms that the transition temperature at fixed chemical potential $\mu < \mu^c_{T = 0}$ ($\mu > \mu^c_{T = 0}$ grows as $\nu$ is increased (decreased), asymptoting to infinity when $\nu \to \nu_c = 1/\sqrt{3}$. When $\mu < \mu^c_{T = 0} \simeq 305$ MeV, the chirally broken phase lies under the transition lines and to the right of the vertical asymptote at $\nu = \nu_c$. When $\mu > \mu^c_{T = 0}$, the chirally restored phase lies under the transition line (extending to the right) and the system above this line (including the whole area to the left of the vertical asymptote $\nu = \nu_c$) lies in the chirally broken phase. These results are consistent with the conjectured phase diagram shown in Fig.~4 of Ref.~\cite{Chen:2022smf}, in the context of the deconfinement transition in pure Yang-Mills, though in that case, the critical rotation parameter is $\nu_c^{\rm YM} = 1/4$.

Panel (b) of Fig.~\ref{fig:pd_axis} shows the phase diagram in the $T$-$\mu$ plane for various values of $\nu$. As $\nu$ is increased, the transition line migrates monotonically upward (there are no intersections between the transition lines corresponding to different values of $\nu$). As $\nu$ approaches $\nu_c = 1/\sqrt{3}$, the transition line continues to move upward. For $\nu > \nu_c$, the region on the lower left part of the diagram remains in the chirally broken phase and the transition line bends towards the right. In this case, the chirally restored phase resides in the bottom right part of the diagram and the transition line keeps traveling toward the lower right corner as $\nu$ is increased. The chirally restored phase reaches its smallest extent when $\nu = 1$. As in the far-field case discussed in the previous section, the transition at $T = 0$ is independent of the value of $\nu$, therefore all transition lines meet at the point $(T, \mu) = (0, \mu^c_{T = 0})$ with $\mu^c_{T = 0} \simeq 305$ MeV.

\subsection{Finite distances}\label{sec:pd:l}

\begin{figure}[t]
    \centering
    \includegraphics[width=.95\linewidth]{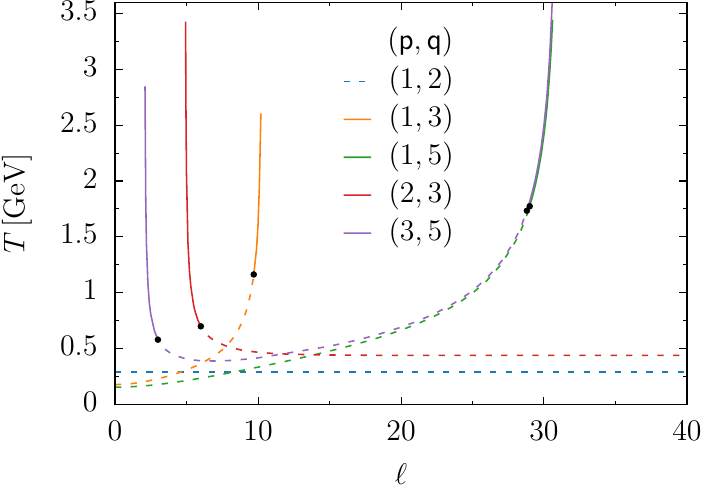} \vspace{-160pt} \\ 
\hspace{-.55\linewidth} (a) \vspace{150pt}\\
    \includegraphics[width=.95\linewidth]{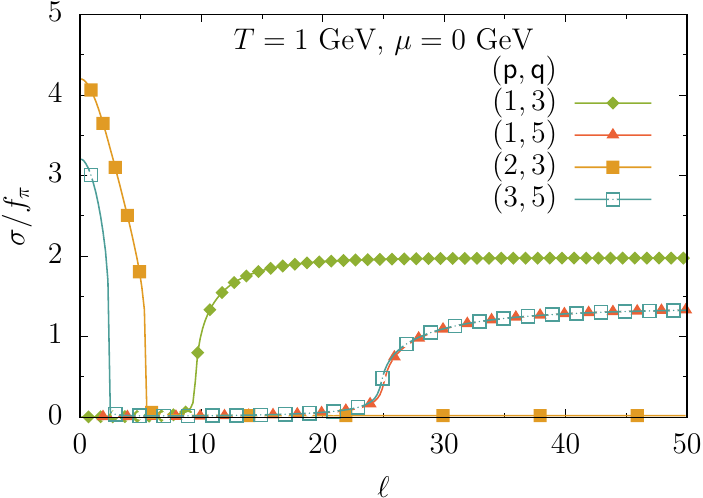} \vspace{-160pt} \\ 
\hspace{-.65\linewidth} (b) \vspace{150pt}
    \caption{(Top) Phase diagram in the $T$-$\ell$ plane at $\mu = 0$ and various values of $\nu = \mathsf{p} / \mathsf{q}$. (Bottom) Dimensionless meson condensate $\sigma / f_\pi$ as a function of $\ell$ for $T = 1$ GeV, $\mu =0$ and various values of $\nu = \mathsf{p} / \mathsf{q}$}
    \label{fig:TvsEllMu0}
\end{figure}

At finite dimensionless distance $\ell = 2\pi \rho T$ from the rotation axis, the system transitions between the $\ell = 0$ and $\ell = \infty$ phase diagrams. We first discuss the dependence of the transition temperature on $\ell$ at vanishing chemical potential, for a few selected values of $\nu = \mathsf{p}/\mathsf{q}$, shown in Fig.~\ref{fig:TvsEllMu0}. First, we note that for $\nu = 1/2$, the thermodynamic potential and the fermion condensate are independent of $\ell$, as discussed in Sec.~\ref{sec:rot:particular}. Therefore, the phase of the system is completely homogeneous in this case and the transition temperature is just a horizontal line (shown with blue color). For $\nu = 1/3$ and $1/5$, the parameter $\mathsf{k} = \mathsf{p} + \mathsf{q}$ is even and, according to Fig.~\ref{fig:pd_farfield}, the system at large distances resides in the broken phase for all temperatures, when $\mu < \mu^c_{T = 0} \simeq 305$ MeV. However, on the rotation axis, since $\nu < \nu_c = 1/\sqrt{3}$, the transition temperature is finite. The transition between these two limits is achieved by the divergence of the transition temperature at a finite distance $\ell$ around $10$ for $\nu = 1/3$ and $30$ for $\nu = 1/5$, as indicated by the orange and green lines, respectively. In the case $\nu = 2/3$, $\mathsf{k}$ is odd and the transition in the far-field limit occurs at $T_{c;\mathsf{q}} = \mathsf{q} T_{c;{\rm static}} \simeq 440$ MeV. However, since $\nu > \nu_c$, the system resides in the broken phase on the rotation axis for any value of the temperature, as long as $\mu < \mu^c_{T = 0}$. Therefore, in this case, the transition between the two regimes occurs in the opposite direction, with the critical temperature diverging as $\ell$ is decreased toward a critical value of around $\ell = 5$ (see red line). Consider now the case $\nu = 3/5$, which exceeds $\nu_c$ and for which $\mathsf{k}$ is even. Thus, the system resides in the broken phase both on the rotation axis and in the far-field limit. There is an intermediate region between $1 \lesssim \ell \lesssim 30$ where the system can reside in the restored phase for high enough temperatures. Therefore, when $T \gtrsim 450$ MeV, we encounter an exotic inhomogeneous phase, with a chirally broken core close to the rotation axis, surrounded by a chirally restored ring at intermediate distances, and finally a chirally broken phase at large distances. While the left asymptote at $\ell \simeq 1$ is located before the one for $\nu = 2/3$, the right asymptote seems to coincide with the one for $\nu = 1/5$, which suggests the interesting conjecture that for $\mathsf{k}$ even, the critical distance at which the system enters the broken phase regardless of temperature is a function only of the denominator $\mathsf{q}$ ($\mathsf{q} = 5$ in this case). 

Panel (b) of Fig.~\ref{fig:TvsEllMu0} illustrates the behavior of the $\sigma$ condensate as a function of $\ell$, at $T = 1$ GeV and $\mu = 0$, for $\nu = 1/3$, $1/5$, $2/3$ and $3/5$. In the first two cases, the system is in the restored phase close to the rotation axis and in the broken phase in the far-field region; in the third case, the situation is reversed: the system is in the chirally broken phase close to the rotation axis and in the restored phase in the far-field limit; in the fourth case, the system transitions from the broken phase close to the rotation axis to the restored phase at intermediate distances and back to the broken phase in the far-field limit. We remark that the exotic scenario corresponding to $\nu > \nu_c$ and odd $\mathsf{k}$ gives the spatial phase ordering observed in lQCD simulations \cite{Braguta:2023yjn,Braguta:2023tqz,Braguta:2024zpi,Yang:2023vsw}.

\begin{figure}[t]
    \centering
    \begin{tabular}{c}
    \includegraphics[width=.95\columnwidth]{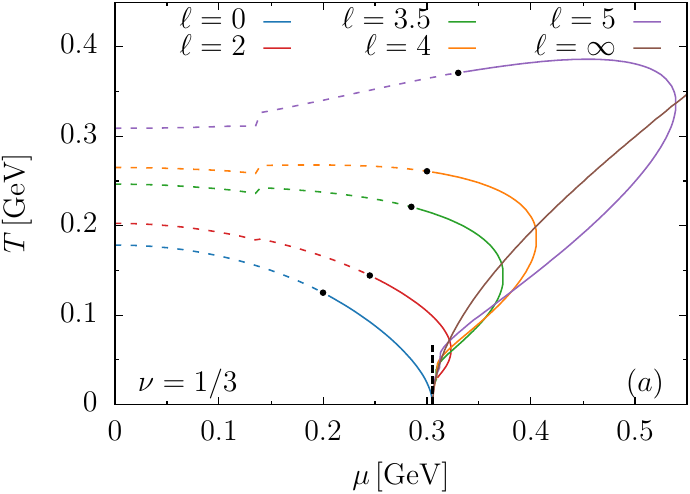} \\
    \includegraphics[width=.95\columnwidth]{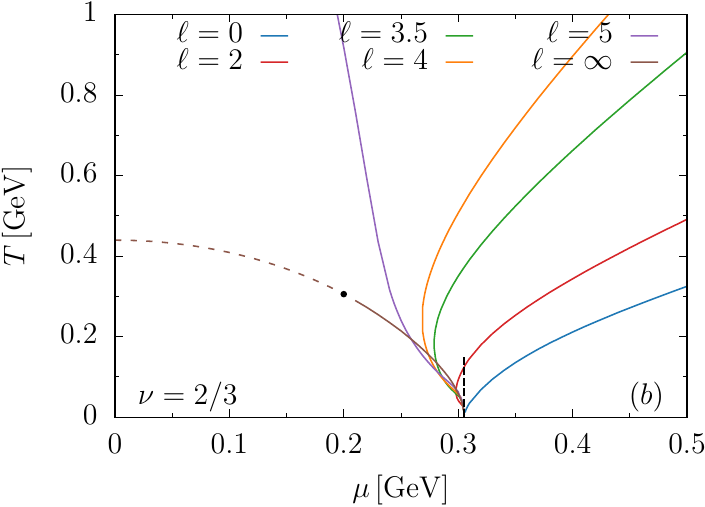}
    \end{tabular}
    \caption{Phase diagram in the $T$-$\mu$ plane, for dimensionless imaginary rotation parameter $\nu = \Omega_I / 2\pi T = 1/3$ (a) and $2/3$ (b). Each curve corresponds to a different value of the dimensionless distance $\ell = 2\pi \rho T$.
    }
    \label{fig:pd_finite}
\end{figure}

We now discuss the phase diagram of the system at intermediate distances from the rotation axis. Figure~\ref{fig:pd_finite} shows the transition lines in the $T$-$\mu$ plane for various values of $\ell$, in the cases $\nu = 1/3$ (a) and $2/3$ (b). 

In the first case, shown in Fig.~\ref{fig:pd_finite}(a), $\nu = 1/3 < 1/\sqrt{3}$ and the phase diagram on the rotation axis ($\ell = 0$) closes at $\mu = 0$, when $T_c \simeq 178$ MeV. Since $\mathsf{k} = 1 + 3$ is even, the role of the temperature is reversed at large distances from the rotation axis and the system at high temperatures resides in the chirally broken phase. The transition line for $\ell \to \infty$ is bent toward the right (large $\mu$). The intermediate values of $\ell$ clearly interpolate between these two limit cases, with two features standing out: the transition temperature at vanishing chemical potential monotonically increases with $\ell$; and for any $\ell > 0$, the transition line emanating from the horizontal axis at $T = 0$ bends to the right, towards higher $\mu$. A small jump can be seen in the crossover phase for the intermediate values $\ell = 3.5$, $4$ and $5$, which is due to the fact that the gradient $\partial \sigma / \partial T$ develops two competing local maxima. 

Panel (b) of Fig.~\ref{fig:pd_finite} shows the case $\nu = 2/3$, above the critical value $\nu_c = 1/\sqrt{3}$ and with odd $\mathsf{k} = 2 + 3$. Therefore, at $\ell = 0$, the transition line is bent toward the right side of the diagram, while at $\ell \to \infty$, the transition line is bent toward the left, closing on the $\mu = 0$ axis at $T = 3 T_{c;{\rm static}} \simeq 440$ MeV. The transition lines at intermediate values of $\ell$ interpolate between these limits with the lines for $\ell \lesssim 4$ bent toward the right and those with $\ell \gtrsim 5$ bent toward the left.

\begin{figure}[t]
\centering
\includegraphics[width=.95\linewidth]{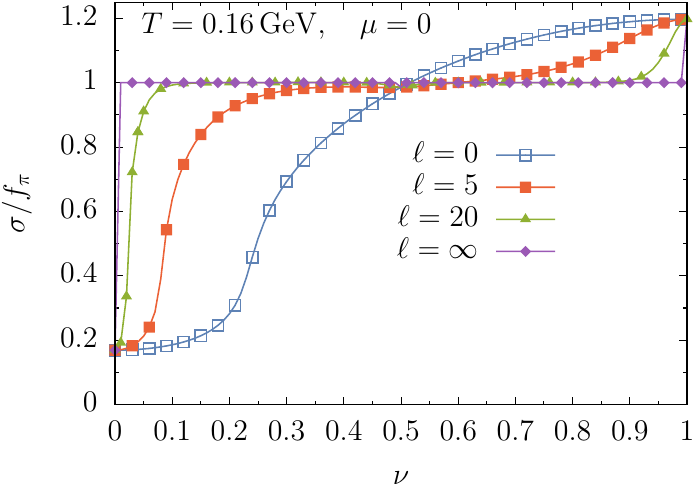}
\caption{
Dimensionless meson condensate $\sigma / f_\pi$ as a function of $\nu$ for various values of $\ell$. The temperature and chemical potential are set to $T = 0.16$ GeV and $\mu = 0$, respectively.
\label{fig:sigma_vs_nu_finitel}
}
\end{figure}

Before ending this section, we discuss the effect of the rotation parameter $\nu = \beta \Omega_I / 2\pi$ on the meson condensate $\sigma$. Figure~\ref{fig:sigma_vs_nu_finitel} shows $\sigma / f_\pi$ as a function of $\nu$ for $\ell = 0$, $5$, $20$ and the far-field limit ($\ell \to \infty$). We set the temperature to $T = 160$ MeV, above the transition temperature in the static limit ($\nu= 0$), and considered for simplicity the case of vanishing chemical potential, $\mu = 0$. On the rotation axis, it can be seen that increasing $\nu$ leads to the increase of $\sigma$, with a cross-over phase transition to the chirally broken phase around $\nu \simeq 0.25$, in agreement with the results in Fig.~\ref{fig:pd_axis}. As $\ell$ is increased, the transition occurs at smaller values of $\nu$, as suggested also by Fig.~\ref{fig:TvsEllMu0}. In the far-field limit, $\sigma = f_\pi$ for all considered values of $\nu$, except the two limiting cases $\nu = 0$ and $\nu = 1$, corresponding to $\mathsf{q} = 1$, with odd and even $\mathsf{k}$, respectively. In the former case, we have the standard fermionic behavior of the static system, while in the latter case, the fermion condensate exhibits a ghost-like behavior and the nonvanishing temperature $T_{\mathsf{q}} = T = 160$ MeV leads to an effective increase of $\sigma$. The results shown in Fig.~\ref{fig:sigma_vs_nu_finitel} are in qualitative agreement with those obtained using the functional renormalization group technique applied to the quark-meson model in Ref.~\cite{Chen:2023cjt} (see Fig.~9 therein for their result for imaginary rotation).

\begin{figure}[t]
    \centering
\begin{tabular}{c}
    \includegraphics[width=.95\linewidth]{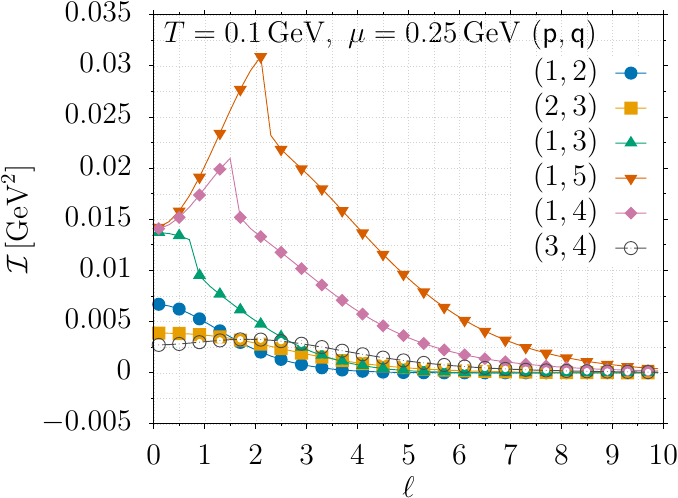} \vspace{-40pt} \\ 
\hspace{.8\linewidth} (a) \vspace{28pt}\\
    \includegraphics[width=.95\linewidth]{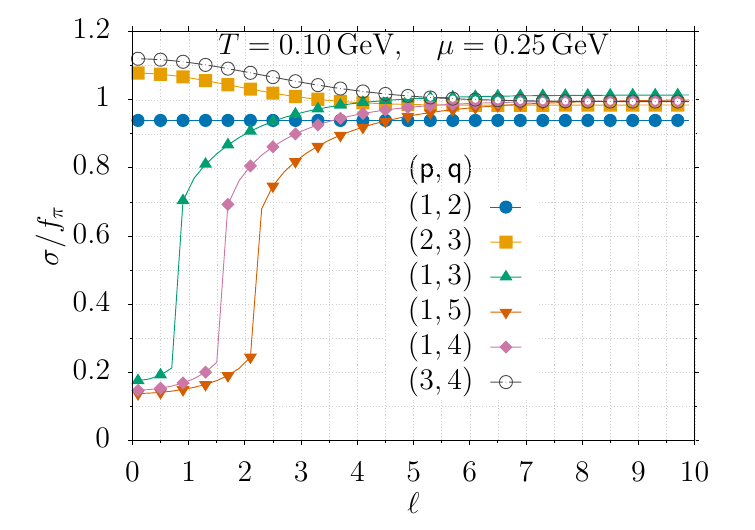} \vspace{-42pt} \\ 
\hspace{.7\linewidth} (b) \vspace{30pt}
\end{tabular}
    \caption{Moment of inertia $\mathcal{I}$ (top) and dimensionless meson condensate $\sigma / f_\pi$ (bottom) as functions of the dimensionless distance $\ell = 2 \pi \rho T$ to the rotation axis, for various values of the rotation parameters $\nu = \beta \Omega_I /2\pi = \mathsf{p} / \mathsf{q}$. The temperature and chemical potential on the rotation axis are set to $T = 0.1$ GeV and $\mu = 0.25$ GeV, respectively.}
    \label{fig:I_p_over_q}
\end{figure}

\section{Moment of inertia}\label{sec:I}

We now consider the moment of inertia density, $\mathcal{I} = -\mathcal{M}_I / \Omega_I$, as a function of the dimensionless radial distance $\ell = 2\pi \rho / \beta$, in the frame of the linear sigma model under imaginary rotation.

\subsection{Generic features} \label{sec:I:generic}

Figure~\ref{fig:I_p_over_q}(a), showing the typical behavior of $\mathcal{I}$, calculated for $(T,\mu) = (0.1\ {\rm GeV}, 0.25\ {\rm GeV})$, reveals several features, which we discuss below, taking into account the phase in which the system resides, indicated by the values of the meson condensate $\sigma / f_\pi$, shown in Fig.~\ref{fig:I_p_over_q}(b). 

For $\nu = 1/2$, $2/3$ and $3/4$, $\mathsf{k}$ is odd and the system resides in the chirally broken phase throughout the domain ($\sigma \gtrsim f_\pi$). While for $\nu = 1/2$, the thermodynamic potential and the fermion condensate are independent of $\ell$, the moment of inertia density exhibits a decrease from a finite value on the rotation axis toward $0$ as $\ell$ is increased, as discussed in Sec.~\ref{sec:rot:particular}. Therefore, we have the counterintuitive result that the total moment of inertia of the system enclosed inside a cylinder of radius $R$ and height $H$ tends to an asymptotic finite value as $R \to \infty$, obtained by integrating Eq.~\eqref{eq:MI_1_over_2}:
\begin{align}
 I\rvert_{\substack{\mathsf{p} = 1\\ \mathsf{q} = 2}}(R) &= -\frac{1}{\pi T} \int_{\rho < R} d^3x\,\mathcal{M}_I\rvert_{\substack{\mathsf{p} = 1 \\ \mathsf{q} = 2}} \nonumber\\
 &= \frac{N_c N_f H}{4 \pi^2 T} \sum_{\varsigma = \pm 1} \int_M^\infty \frac{dE\, E}{p \cosh(\beta \mathcal{E}_\varsigma)} \sin^2(p R).
\end{align}
While for $\nu = 2/3$, the moment of inertia exhibits a similar monotonic decrease with increasing $\ell$, for $\nu = 3/4$ we observe a slight increase up to $\ell \simeq 2.5$. 

Moving on to the cases $\nu = 1/3$, $1/4$ and $1/5$, Fig.~\ref{fig:I_p_over_q}(b) shows that the system is in the chirally restored phase close to the rotation axis, transitioning toward the chirally broken phase as $\ell$ is increased. At the point of transition, $\mathcal{I}$ exhibits a sudden drop in magnitude. While for $\nu = 1/3$, $\mathcal{I}$ is a monotonically decreasing function of $\ell$, in the cases $\nu = 1/4$ and $1/5$ we observe an increase in $\mathcal{I}$ until the transition point, followed by a monotonic decrease as $\ell$ is further increased. We note that the position of the peak (corresponding to the transition point) shifts to higher $\ell$ and its magnitude increases as $\nu$ decreases, a feature which will be discussed in the next paragraph.

\begin{figure}
\centering
\begin{tabular}{c}
     \includegraphics[width=.95\columnwidth]{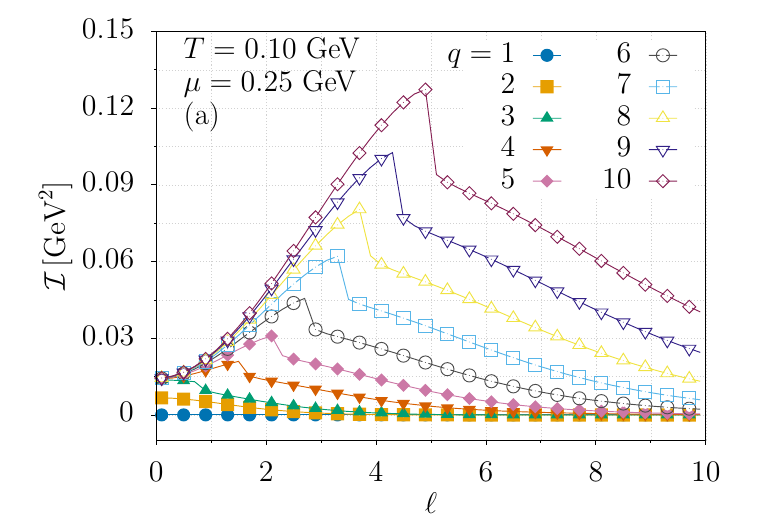}  \\
     \includegraphics[width=.95\columnwidth]{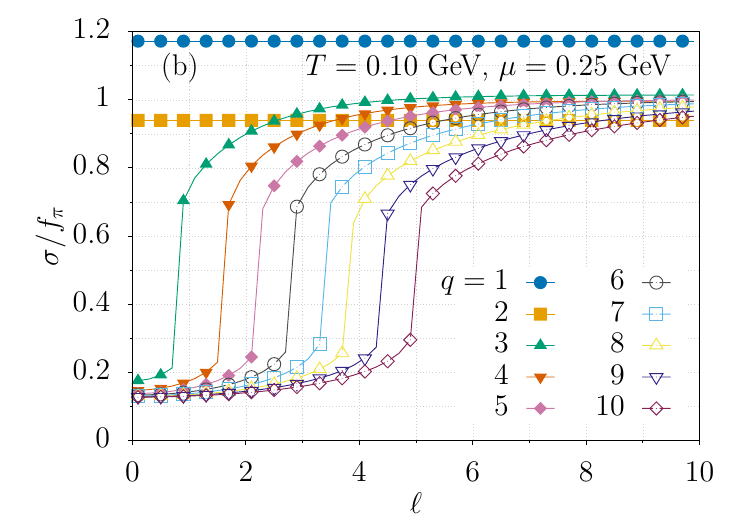}  
\end{tabular}
\caption{Moment of inertia $\mathcal{I}$ (top) and dimensionless meson condensate $\sigma / f_\pi$ (bottom) as functions of the dimensionless distance $\ell = 2 \pi \rho T$ to the rotation axis, for rotation parameters $\nu = \beta \Omega_I /2\pi = 1/\mathsf{q}$, with $1 \le \mathsf{q} \le 10$. The temperature and chemical potential on the rotation axis are set to $T = 0.1$ GeV and $\mu = 0.25$ GeV, respectively.
\label{fig:I_1_over_q}
}
\end{figure}

In Fig.~\ref{fig:I_1_over_q}, we analyze the properties of $\mathcal{I}$ for decreasing $\nu$, taken as $\nu = 1/\mathsf{q}$, with $\mathsf{q}$ between $1$ and $10$. As already established in Fig.~\ref{fig:I_p_over_q}, $\mathcal{I}$ exhibits an increase with $\ell$ until the transition point. This figure confirms that the height of the peak increases as $\nu$ is decreased. We can better understand the features shown in this plot by looking at the moment of inertia for slow rotation, discussed in the next paragraph.

\subsection{Vanishing rotation limit}\label{sec:I:Omega0}

For $\nu = 1/\mathsf{q} \to 0$ as $\mathsf{q} \to \infty$, we compare the result in Eq.~\eqref{eq:sumr_thermo} to the $\Omega_I \to 0$ limit of $-\Delta \mathcal{M}_I / \Omega_I$ taken in Eq.~\eqref{eq:thermodynamic_vsum}. Using 
\begin{equation}
 \frac{1}{4\rho} \frac{d}{d\rho} \frac{1}{\rho} \frac{d}{dT} \left[\frac{c_v}{s_v^3 \Omega_I^2} \sin(2 p \rho s_v)\right] \simeq -\frac{v^2 p^3}{30T^3}(5 + 4 p^2 \rho^2),
\end{equation}
as well as the summation formula
\begin{equation}
 \sum_{v = 1}^\infty (-1)^{v+1} v^2 {\rm sgn}(\mathcal{E}_\varsigma) e^{-v \beta |\mathcal{E}_\varsigma|} = T^2 \frac{d^2}{dE^2} \left(\frac{1}{e^{\beta \mathcal{E}_\varsigma} + 1}\right),
\end{equation}
we arrive at
\begin{align}
 \mathcal{I}\rvert_{\Omega_I = 0} &= \frac{N_c N_f}{60\pi^2} \sum_{\varsigma = \pm 1} \int_M^\infty dE\, p^3(5 + 4\rho^2 p^2) \frac{d^2}{dE^2} \frac{1}{e^{\beta \mathcal{E}_\varsigma} + 1} \nonumber\\
 &= 2 N_c N_f \sum_{\varsigma = \pm 1} \int \frac{dP}{e^{\beta \mathcal{E}_\varsigma} + 1} \left[\frac{1}{4} \left(1 + \frac{E^2}{p^2}\right) \right.\nonumber\\
 &\left. + \rho^2 \left(E^2 + \frac{p^2}{3}\right)\right],
 \label{eq:I_Omega0}
\end{align}
where $dP = d^3p / [(2\pi)^3 E]$, in agreement with Eq.~(99) in Ref.~\cite{Patuleanu:2025zbn}. 

The second term in Eq.~\eqref{eq:I_Omega0} corresponds to the product $\rho^2 h$ between the squared distance to the rotation axis and the enthalpy density, $h = \epsilon + P$, with the energy density and pressure of the Fermi gas given by
\begin{equation}
 \begin{pmatrix}
  \epsilon \\ P
 \end{pmatrix} = 2N_c N_f \sum_{\varsigma = \pm 1} \int \frac{dP}{e^{\beta \mathcal{E}_\varsigma} + 1} \begin{pmatrix}
     E^2 \\ p^2 / 3
 \end{pmatrix}.
\end{equation}
This term therefore represents the classical, orbital contribution to the moment of inertia, being the relativistic generalization of the Newtonian formula $\rho^2 \varrho$, where $\varrho$ is the mass density. 

Separating $\mathcal{I}\rvert_{\Omega_I = 0}= \mathcal{I}_L + \mathcal{I}_\Sigma$, with $\mathcal{I}_L = \rho^2 h$ being the orbital (classical) contribution, we recognize that the coordinate-independent spin contribution $\mathcal{I}_\Sigma = \frac{1}{2} \sigma^\omega_A\rvert_{\Omega = 0}$, given by the first term in Eq.~\eqref{eq:I_Omega0}, is equal to half of the axial vortical conductivity at vanishing rotation, $\sigma^\omega_A = J^z_A / \Omega$, defined as the ratio between the component $J^z_A$ of the axial current $J^\mu_A = \bar{\psi} \gamma^\mu \gamma^5 \psi$ along the axis of rotation and the rotation angular velocity $\Omega$. This quantity can be expressed as (see Eq.~(7.4) of Ref.~\cite{Ambrus:2019ayb}):
\begin{equation}
 \sigma^\omega_A\rvert_{\Omega = 0} = N_c N_f \sum_{\varsigma = \pm 1} \int \frac{dP}{e^{\beta \mathcal{E}_\varsigma} + 1} \left(1 + \frac{E^2}{p^2}\right).
\end{equation}

\begin{figure}
    \centering
    \includegraphics[width=\linewidth]{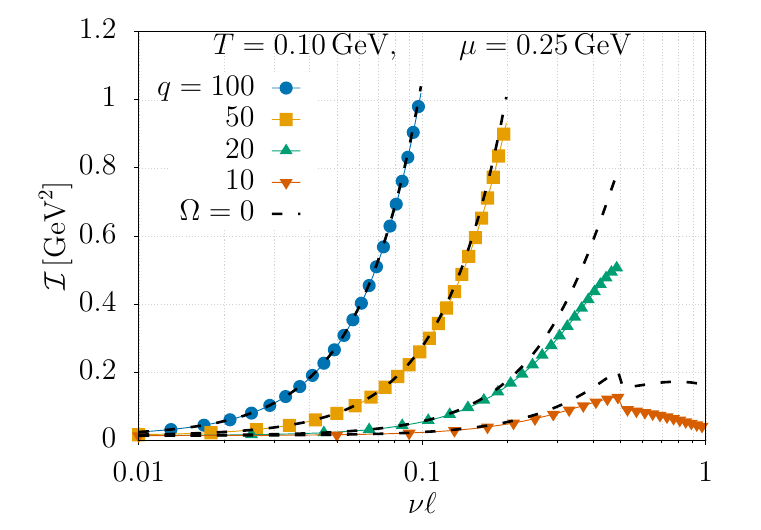}
    \caption{Comparison between the moment of inertia density computed for $\nu = 1 / \mathsf{q}$, with $\mathsf{q} = 100$, $50$, $20$ and $10$, and the expression \eqref{eq:I_Omega0} at vanishing rotation, evaluated using the local values of the fermion mass $M = g\sigma$, temperature $T_\rho$ and chemical potential $\mu_\rho$ [see Eq.~\eqref{eq:Tmu_local}], shown with respect to $\nu \ell = \rho \Omega_I$ in logarithmic scale.}
    \label{fig:I_Omega0}
\end{figure}

To facilitate the comparison between the moment of inertia in the vanishing rotation limit, given in Eq.~\eqref{eq:I_Omega0}, and the results obtained at finite $\nu = 1/\mathsf{q}$, we acknowledge that in the system under rotation, the local fermion mass $M = g \sigma$ is point-dependent, while the local temperature and chemical potential are given by the Tolman-Ehrenfest formula \eqref{eq:Trho},
\begin{equation}
 T_\rho = \Gamma_\rho T, \quad \mu_\rho = \Gamma_\rho \mu, \label{eq:Tmu_local}
\end{equation}
with $\Gamma_\rho = 1/\sqrt{1 + \rho^2 \Omega_I^2} = 1 / \sqrt{1 + \ell^2 \nu^2}$ the local Lorentz factor. 

Figure~\ref{fig:I_Omega0} represents $\mathcal{I}$ with respect to the product $\nu \ell$, in logarithmic scale, for $\nu = 1/ \mathsf{q}$ with $\mathsf{q} = 10$, $20$, $50$ and $100$. It can be seen that the expression \eqref{eq:I_Omega0} at vanishing rotation (shown with dotted black lines) provides a reasonable approximation up to $\nu \ell \simeq 0.2$. For large $\mathsf{q}$ (small $\nu = 1/\mathsf{q}$), the system remains in the chirally restored phase and $\mathcal{I}$ increases approximately quadratically with $\ell$ over the represented domain. The curve corresponding to $\mathsf{q} = 10$ exhibits a sharp drop due to the transition to the chirally broken phase around $\nu \ell = 0.5$, visible also in the approximation corresponding to vanishing rotation, however the two lines remain in quantitative disagreement, signaling the breakdown of the vanishing rotation approximation at large values of $\nu \ell$.

\subsection{Large chemical potential}\label{sec:I:largemu}

\begin{figure}
\begin{tabular}{c}
\includegraphics[width=.95\linewidth]{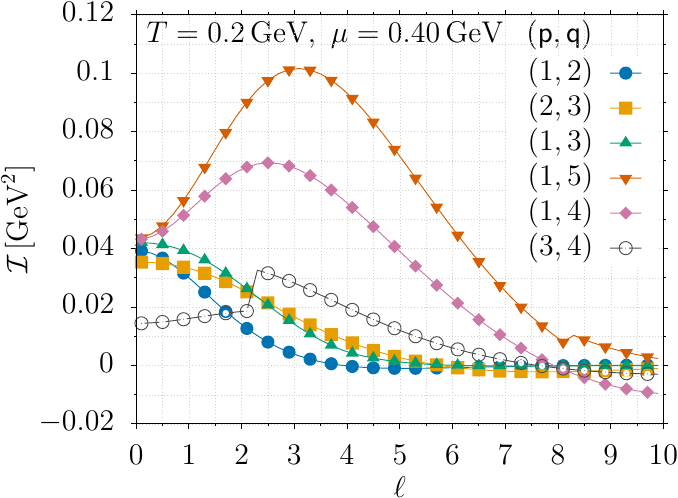} \\
\includegraphics[width=.95\linewidth]{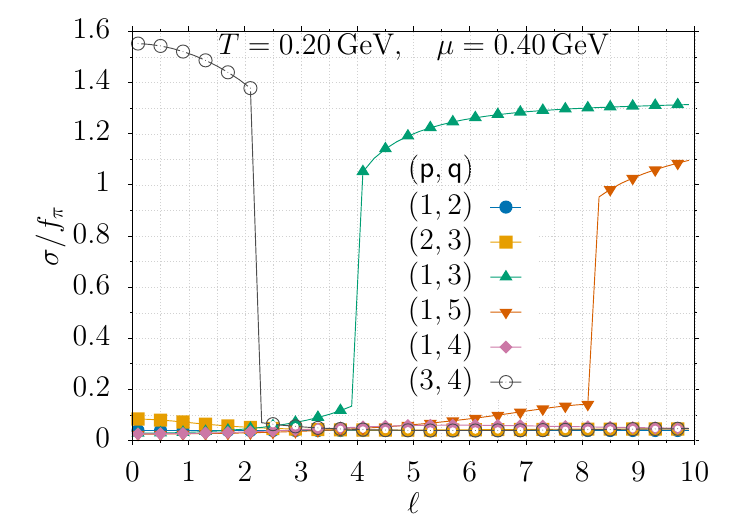}
\end{tabular}
\caption{Moment of inertia $\mathcal{I}$ (top) and dimensionless meson condensate $\sigma / f_\pi$ (bottom) as functions of $\ell = 2 \pi \rho T$, for various values of the rotation parameter $\nu = \mathsf{p} / \mathsf{q}$, computed at $T = 0.2$ GeV and $\mu = 0.4$ GeV.
\label{fig:I_largemu}
}
\end{figure}

We now consider the case of high chemical potential. Considering the expression \eqref{eq:massless_sumv} for the non-degenerate contribution $\Delta \phi$ to the thermodynamic potential at vanishing mass, we see that the oscillatory term $c^{v\mu} = \cos(2\rho \mu s_v)$ is amplified when $\mu$ is large, potentially leading to a sign change with respect to $\rho$ before the moment of inertia density $\mathcal{I} = \frac{1}{\Omega_I} \partial \Delta \phi / \partial \Omega_I$ becomes negligible. 

Figure~\ref{fig:I_largemu} shows the moment of inertia density (top panel) and $\sigma$ condensate (bottom panel) for $T = 0.2$ GeV and $\mu = 0.4$ GeV, for various values of $\nu = \mathsf{p} / \mathsf{q}$, with respect to $0 \le \ell \le 10$. Looking at the bottom panel, we see that for $\nu = 3/4 > \nu_c$ and odd $\mathsf{k} = 3 + 4 = 7$, the system transitions from the chirally broken phase on the rotation axis to the restored phase around $\ell \simeq 2.25$. At the transition point, $\mathcal{I}$ exhibits a jump. For $\nu = 1/3$ and $1/5$, which are smaller than $\nu_c$ but correspond to even $\mathsf{k}$, the system transitions from the chirally restored phase on the rotation axis to the broken phase around $\ell \simeq 4$ and $8.25$, respectively, however no notable jump can be seen in $\mathcal{I}$. For $\nu = 1/2$ and $1/4$, the system resides in the chirally restored phase for all values of $\ell$. 

In all cases, we see that $\mathcal{I}$ is positive close to the rotation axis. A dip to negative values can be seen in the cases $\nu = 3/4$ and $1/4$, around $\ell \gtrsim 8$, with the curve for $\nu= 1/4$ achieving a more negative value than $\nu = 3/4$. In both cases, the negative contributions to $\mathcal{I}$ come from outside the causal cylinder, bounded by $\nu \ell = 1$. 

\section{Conclusion}\label{sec:conc}

In this paper, we considered the properties of strongly interacting matter at finite temperature $T = \beta^{-1}$ and chemical potential $\mu$, undergoing rigid rotation with an imaginary, constant angular velocity, $\Omega = i \Omega_I$ (with $\Omega_I$ a real number). Our study employed the linear sigma model coupled with quarks to model the QCD chiral phase transition.

Owing to the phenomenon of fractalization of thermodynamics, the system develops a peculiar behaviour far from the rotation axis (in the far-field limit). Here, the thermodynamics of the system is governed by the denominator $\mathsf{q}$ in the irreducible fraction representation of the dimensionless rotation parameter, $\nu = \beta \Omega_I / 2\pi = \mathsf{p} / \mathsf{q}$, and on whether the number $\mathsf{k} = \mathsf{p} + \mathsf{q}$ is odd or even. Specifically, for odd $\mathsf{k}$, the system resembles a static system with the temperature $T_{\mathsf{q}} = T / \mathsf{q}$ reduced by a factor of $\mathsf{q}$, while the chemical potential remains unchanged. We note that the system does not achieve thermal equilibrium, as the entropy is a factor of $\mathsf{q}$ smaller than that corresponding to the static system with temperature $T_{\mathsf{q}}$. When $\mathsf{k}$ is even, the statistics of the system shifts from fermionic to ghost-like, with the distribution function becoming the Bose-Einstein distribution with an overall negative sign. 

Our investigations reveal an immediate impact on the phase diagram of the system: in the odd-$\mathsf{k}$ case, the transition line in the $T$-$\mu$ plane can be obtained from that of the nonrotating (static) system by multiplying the temperature axis by $\mathsf{q}$, such that the transition temperature at a given value of $\mu$ becomes $\mathsf{q}$ times larger than in the static system. When $\mathsf{k}$ is even, the ghost-like behavior affects the thermal contributions to the thermodynamic potential, modifying their sign. Therefore, increasing the temperature has the effect of inhibiting the chiral phase transition. Therefore, the system resides in the chirally broken phase for all temperatures, when the chemical potential is less than the value $\mu^c_{T = 0} \simeq 305$ MeV corresponding to the critical value of the chemical potential in the static system at vanishing temperature. When $\mu > \mu^c_{T = 0}$, the system is chirally restored at low temperatures, transitioning to the chirally broken phase when $T$ is increased.

We further investigated the phase diagram of the system on the rotation axis. Here, the thermodynamic potential receives corrections due to rotation. At the level of the fermion condensate for negligible fermion mass (i.e., in the chirally restored phase), we identified a critical value $\nu_c = 1/\sqrt{3}$ of the rotation parameter for which the sign of the temperature-dependent term is flipped. For $\nu < \nu_c$, the transition temperature increases with increasing $\nu$, at fixed chemical potential $\mu < \mu^c_{T = 0}$, the system being in the restored phase for any temperature when $\mu > \mu^c_{T = 0}$. On the contrary, when $\nu > \nu_c$, increasing the temperature inhibits chiral restoration, such that the system resides in the broken phase for all values of $T$ when $\mu < \mu^c_{T = 0}$, while for $\mu > \mu^c_{T = 0}$, the chirally restored phase is confined to small values of $T$. 

At finite distances from the rotation axis, expressed through the dimensionless parameter $\ell = 2\pi \rho T$, the shape of the transition line interpolates between the $\rho = 0$ case and the far-field ($\rho \to \infty$) case. First, we illustrated in Fig.~\ref{fig:TvsEllMu0} how the transition temperature at vanishing chemical potential depends on $\ell$, identifying three regimes with inhomogeneous phases: (1) finite at $\ell = 0$ and diverging at a finite $\ell$, leading to a chirally broken state in the far-field limit for any temperature ($\nu < \nu_c = 1/\sqrt{3}$ and even $\mathsf{k} = \mathsf{p} + \mathsf{q}$); (2) finite in the far-field limit and diverging as $\ell$ is decreased to a non-vanishing value, leading to a chirally broken phase in the vicinity of the rotation axis for any temperature ($\nu > \nu_c$ and odd $\mathsf{k}$); and (3) chirally broken both in the vicinity of the rotation axis and in the far-field limit, with an intermediate region where chiral restoration is possible ($\nu > 1/\sqrt{3}$ and even $\mathsf{k} = \mathsf{p} + \mathsf{q}$). Out of these, scenario (2) gives the inhomogeneous phase structure observed in lQCD calculations.

We then explored in Fig.~\ref{fig:pd_finite} how the transition line in the $T$-$\mu$ plane migrates with increasing $\ell$ between the rotation axis and far-field behaviors. Panel (a), considering $\nu = 1/3 < \nu_c$ with even $\mathsf{k}$, shows how the transition line is ``flung'' from closing on the $\mu = 0$ axis for $\ell = 0$ to bending towards the right side of the diagram in the far-field limit; while panel (b), with $\nu = 2/3 > \nu_c$ and odd $\mathsf{k}$, displays the opposite behavior: the transition line is ``flung'' from closing on the $\mu = 0$ axis in the far-field limit to bending towards the right side of the diagram on the rotation axis. 

We finally considered the behavior of the moment of inertia density $\mathcal{I}$ of the system, discussed in Sec.~\ref{sec:I}. First, we showed that on the rotation axis, the moment of inertia is always positive. Figure~\ref{fig:I_Omega0} confirms that as $\nu = 1/\mathsf{q}$ decreases, we see agreement with the moment of inertia computed in the limit of vanishing (real or imaginary) rotation, only up to $\nu \ell = \rho \Omega_I \simeq 0.2$. Typically, $\mathcal{I} = \mathcal{I}_L + \mathcal{I}_\Sigma$ is comprised of a local spin contribution $\mathcal{I}_\Sigma = \frac{1}{2} \sigma^\omega_A$, related to the axial vortical conductivity $\sigma^\omega_A$, and an orbital contribution $\mathcal{I}_L = \rho^2 h$, with $h = \epsilon + P$ being the enthalpy density. At small $\nu$, $\mathcal{I}$ exhibits a clear quadratic increase with respect to $\ell$, confirmed in Fig.~\ref{fig:I_1_over_q}, with a sharp drop occurring at the point where the system transitions from the chirally restored to the chirally broken phase. At large $\ell$, $\mathcal{I}$ monotonically approaches $0$ for any non-vanishing value of $\nu$, thus breaking the analogy with the system under real rotation. The onset of this behavior takes place around $\nu \ell \simeq 0.5$, well within the causal cylinder bounded by $\nu \ell = 1$ [see Fig.~\ref{fig:I_1_over_q}(a)].

We also considered the properties of the moment of inertia density $\mathcal{I}$ at large chemical potential, motivated by the oscillatory term $c^{v\mu} = \cos(2\rho \mu s_v)$ appearing in the massless limit of the nondegenerate contribution $\Delta\phi$ to the thermodynamic potential density, shown in Eq.~\eqref{eq:massless_sumv}. Our results in Fig.~\ref{fig:I_largemu} showed that at $\mu = 0.4\ {\rm GeV} > \mu^c_{T = 0} \simeq 0.305$ GeV, the system with $\nu = 1/4$ does indeed achieve a negative moment of inertia for $\ell \gtrsim 8$, which however resides outside the causal cylinder, bounded by $\ell = 1/\nu = 4$. We therefore conclude that the negative moment of inertia reported in lattice QCD calculations does not originate from the fermion sector, thus supporting the conclusion of Ref.~\cite{Braguta:2023yjn,Braguta:2023tqz} that it may originate from the evaporation of the nonperturbative chromomagnetic gluon condensate.

Our findings presented in this paper shed light on the complex interplay of the nonperturbative fractal thermodynamics of quantum systems under imaginary rotation and the chiral symmetry restoration phase transition in QCD.

\section*{Acknowledgements} 
The authors are grateful to Dr. Nyx Shiva, Lehel Csillag and Dr. Maxim Chernodub for useful comments on the manuscript. This work was funded by the EU’s NextGenerationEU instrument through the National Recovery and Resilience Plan of Romania - Pillar III-C9-I8, managed by the Ministry of Research, Innovation and Digitization, within the project entitled ``Facets of Rotating Quark-Gluon Plasma'' (FORQ), contract no.~760079/23.05.2023 code CF 103/15.11.2022. 

\appendix

\section{Cauchy integration for ghost-like contributions}\label{app:Cauchy}

As we have seen in Sec.~\ref{sec:rot:rational}, when the rotation parameter takes the rational value $\nu = \beta \Omega_I / 2\pi = \mathsf{p} / \mathsf{q}$ with even $\mathsf{k} = \mathsf{p} + \mathsf{q}$, the fermion expectation values are described by ghost-like statistics, exhibiting a pole due to the emergence of a Bose-Einstein condensate where the effective energy $\mathcal{E}_\varsigma = E - \varsigma \mu$ vanishes. 

We illustrate the Cauchy integration procedure by considering the fractal contribution to the fermion thermodynamic potential density $\phi_{\mathsf{q}}$, shown in Eq.~\eqref{eq:fractal_terms}. Consider the split of $\phi_{\mathsf{q}} = \phi_{\mathsf{q}}^+ + \phi_{\mathsf{q}}^-$ in the two terms corresponding to $\varsigma = \pm 1$, with
\begin{equation}
 \phi_{\mathsf{q}}^\pm = \frac{N_c N_f}{3\pi^2} \int_M^\infty \frac{dE\, p^3}{e^{\beta_\mathsf{q}(E \mp |\mu|)} - 1}.
\end{equation}
In the term $\phi_{\mathsf{q}}^-$, the exponent in the denominator takes only non-negative values, $E + |\mu|$, and hence it can be obtained straightforwardly using numerical integration.

For the term $\phi_{\mathsf{q}}^+$, the distribution function has a pole when $E = |\mu|$, which occurs whenever $|\mu| > M$.  In this case, the integral is understood in the Cauchy
principal-value sense. To implement this prescription, we split the integration domain into two intervals symmetric about the pole at $E = |\mu|$, namely $M < E < |\mu|$ and $|\mu| < E < 2 |\mu| - M$, together with the regular remainder $ E > 2 |\mu| - M$. The contributions from the two symmetric intervals are then combined according to the Cauchy principal-value prescription, while the remainder is evaluated as an ordinary integral. This can be seen explicitly by introducing the distance $\varepsilon$ from the pole. In the first domain, we switch the integration variable to $\varepsilon = |\mu| - E$; while in the second and third domains, we use $\varepsilon = E - |\mu|$. The integration domain over $\varepsilon$ now coincides between the first two cases, being given by $0 < \varepsilon < |\mu| - M$, while for the third case, we have $\varepsilon > |\mu| - M$. The result can be written as $\phi^+_\mathsf{q} = \phi^c_{\mathsf{q}} + \phi^r_{\mathsf{q}}$, with the Cauchy and remainder parts given by
\begin{align}
 \phi^c_\mathsf{q} &= \frac{N_c N_f}{3\pi^2} \int_0^{|\mu| - M} d\varepsilon \left(\frac{p_+^3}{e^{\beta_\mathsf{q} \varepsilon} - 1} - \frac{p_-^3}{1 - e^{-\beta_{\mathsf{q}} \varepsilon}}\right), \nonumber\\
 \phi^r_\mathsf{q} &= \frac{N_c N_f}{3\pi^2} \int_{|\mu| - M}^\infty \frac{d\varepsilon\, p_+^3}{e^{\beta_\mathsf{q} \varepsilon} - 1},
\end{align}
where the momenta obey $p_\pm = \sqrt{(|\mu| \pm \varepsilon)^2 - M^2}$. It is easy to see that the pole around $E = |\mu|$, equivalent to $\varepsilon = 0$, cancels between the two terms, since:
\begin{align}
 \frac{1}{e^{x} - 1} &=\frac{1}{x} - \frac{1}{2} + O(x), &
 \frac{1}{1 - e^{-x}} &\simeq \frac{1}{x} + \frac{1}{2} + O(x).
\end{align}
Around $\varepsilon = 0$, the integrand of $\phi^c_{\mathsf{q}}$ behaves as
\begin{equation}
 \frac{p_+^3}{e^{\beta_\mathsf{q} \varepsilon} - 1} - \frac{p_-^3}{1 - e^{-\beta_{\mathsf{q}} \varepsilon}} \simeq \frac{6|\mu| p_f}{\beta_\mathsf{q}} - p_f^3 + O(\varepsilon^2),
\end{equation}
with $p_f = \sqrt{\mu^2 - M^2}$, such that, combining the contributions from the two domains around the pole, the resulting integrand no longer requires a principal-value prescription. We thus obtain $\phi_{\mathsf{q}}$ as the sum
\begin{equation}
 \phi_{\mathsf{q}} =\phi_{\mathsf{q}}^- + \phi_{\mathsf{q}}^c + \phi_{\mathsf{q}}^r,
\end{equation}
where the integrals in $\phi_{\mathsf{q}}^-$, $\phi_{\mathsf{q}}^c$ and $\phi_{\mathsf{q}}^r$ are all regular.

\bibliography{refs}

\end{document}